\documentclass[a4paper,fleqn,usenatbib]{mnras}

\usepackage{newtxtext,newtxmath}

\usepackage[T1]{fontenc}
\usepackage{ae,aecompl}

\usepackage{graphicx}	
\usepackage{amsmath}	
\usepackage{amssymb}	
\usepackage{arydshln}
\usepackage{times}

\title[Bayesian inference of the IMF in Composite Stellar Populations]{Hierarchical Bayesian inference of the Initial Mass Function in Composite Stellar Populations}

\author[M. Dries et al.]{
M. Dries,$^{1}$\thanks{E-mail: dries@astro.rug.nl}
S. C. Trager,$^{1}$
L.V.E. Koopmans,$^{1}$
G. Popping,$^{2}$
R.S. Somerville$^{3,4}$
\\
$^{1}$Kapteyn Astronomical Institute, University of Groningen, PO BOX 800, 9700 AV Groningen, The Netherlands\\
$^{2}$European Southern Observatory, Karl-Schwarzschild-Strasse 2,  85748, Garching, Germany\\
$^{3}$Department of Physics and Astronomy, Rutgers, The State University of New Jersey, 136 Frelinghuysen Rd, Piscataway, NJ 08854, USA\\
$^{4}$Center for Computational Astrophysics, Flatiron Institute, 162 5th Ave, New York, NY 10010, USA\\
}

\date{Accepted XXX. Received YYY; in original form ZZZ}

\pubyear{2017}

\begin{document}
\label{firstpage}
\pagerange{\pageref{firstpage}--\pageref{lastpage}}
\maketitle

\begin{abstract}
{The initial mass function (IMF) is a key ingredient in many studies of galaxy formation and evolution. Although the IMF is often assumed to be universal, there is continuing evidence that it is not universal. Spectroscopic studies that derive the IMF of the  unresolved stellar populations of a galaxy often assume that this spectrum can be described by a single stellar population (SSP). To alleviate these limitations, in this paper we have developed a unique hierarchical Bayesian framework for modelling composite stellar populations (CSPs). Within this framework we use a parameterized IMF prior to regulate a direct inference of the IMF. We use this new framework to determine the number of SSPs that is required to fit a set of realistic CSP mock spectra. The CSP mock spectra that we use are based on semi-analytic models and have an IMF that varies as a function of stellar velocity dispersion of the galaxy. Our results suggest that using a single SSP biases the determination of the IMF slope to a higher value than the true slope, although the trend with stellar velocity dispersion is overall recovered. If we include more SSPs in the fit, the Bayesian evidence increases significantly and the inferred IMF slopes of our mock spectra converge, within the errors, to their true values. Most of the bias is already removed by using two SSPs instead of one. We show that we can reconstruct the variable IMF of our mock spectra for signal-to-noise ratios exceeding $\sim$75. }
\end{abstract}




\section{Introduction}
The distribution of stellar masses in a stellar population or a galaxy is described by the current mass function (CMF). The CMF is related to the initial mass function (IMF) which describes the distribution of stellar masses at the time that the stars were born. The concept of an IMF was first introduced by \cite{Salpeter1955} who called it the `original mass function'. Based on star counts of resolved stellar populations, the Milky Way (MW) IMF can be described by a broken power law \citep{Kroupa1993, Kroupa2001} or a lognormal distribution extended with a power law for higher masses \citep{Chabrier2003}. Most studies of the MW's IMF in different environments suggest a universal IMF \citep{Bastian2010}. 

More distant galaxies are unresolved and therefore the IMF cannot be determined directly from star counts. Whereas the Galactic IMF is determined for a spiral galaxy, unresolved stellar population studies often focus on early-type (elliptical and lenticular) galaxies, which may not form stars following the MW's IMF. For those galaxies, the IMF may be inferred by modelling their spectra with a stellar population synthesis (SPS) model \citep[e.g.][]{Spinrad1971, Cenarro2003, Conroy2012a}. Recent spectroscopic studies of early-type galaxies (ETGs) suggest that the IMF is not universal and that the relative number of low-mass ($M \lesssim 0.5$ M$_{\odot}$) to high-mass stars is higher in galaxies with a higher mass / velocity dispersion \citep{Conroy2012b, Spiniello2012, Ferreras2013, LaBarbera2013} or metallicity \citep{MartinNavarro2015b}. Within these ETGs there appears to be a radial trend such that the IMF is more bottom-heavy towards the centre \citep{MartinNavarro2015, vanDokkum2016}. Combined studies of population synthesis with dynamics and/or gravitational lensing support that these ETGs have more stellar mass than predicted on the basis of a MW IMF \citep{Treu2010, Graves2010, Cappellari2012, Conroy2012b, Lyubenova2016}. In contrast to these findings, \cite{Smith2015} and \cite{Newman2016} have found a number of lensed ETGs that favour a Kroupa IMF instead of a bottom-heavy IMF.

Inferring the IMF of a galaxy from its spectrum through a SPS model is not straightforward, however. Since \cite{Tinsley1968}, many SPS models \citep[e.g.][]{Bruzual2003, LeBorgne2004, Maraston2005, Conroy2012a, Vazdekis2015} have been developed to infer the properties of a galaxy that are not directly observable. Every SPS model has its own set of ingredients (basically stellar evolution in the form of isochrones, a spectral library and an IMF supplemented with an arbitrary number of model-specific ingredients/parameters and model assumptions) and each of these ingredients is characterized by uncertainties. The ingredients that are or are not included as well as their uncertainties might affect the inferred IMF.

One very noticeable example of such an `ingredient' is that most SPS models assume a parameterized IMF and therefore do not try to determine the IMF shape but simply determine one or two IMF slopes. This in turn controls the ratio between dwarf and giant stars. \cite{Clauwens2016} show that different parameterizations of the IMF can results in similar dwarf-to-giant ratios. More direct approaches to determine the IMF shape have been developed by \cite{Dries2016} and \cite{Conroy2016}. These methods try to determine the contribution of individual mass bins to the integrated spectrum and use this to infer the IMF.

Dwarfs with $M < 0.5$ $\mathrm{M}_{\odot}$ contribute significantly to the stellar mass of a galaxy. The fraction of the stellar mass in stars with $M < 0.5$ M$_{\odot}$ is $\sim$27\% for a Kroupa IMF between 0.1 M$_{\odot}$ and 100 M$_{\odot}$\footnote{In \cite{Dries2016} the fraction of mass in stars with $M < 0.4$ $\mathrm{M}_{\odot}$ is erroneously reported to be 12\% for a Kroupa IMF between 0.1 and 100 $\mathrm{M}_\odot$. The correct number should be 21\%.}. For a Salpeter IMF this fraction increases to $\sim$47\% and for IMFs more bottom-heavy than Salpeter this fraction is even higher. However, the relative contribution of these dwarfs to an optical spectrum is expected to be not more than a few percent for a MW IMF \citep[][DTK16 hereafter]{Dries2016}. Moreover, the spectra of these dwarfs and the spectra of K and M giants look very similar and this makes it difficult to determine dwarf-to-giant ratios. There are, however, some (gravity-sensitive) spectral features that are known to be sensitive to either dwarfs or giants \citep{Wing1969, Faber1980, Gorgas1993, Worthey_et_al1994, Schiavon1997a, Schiavon1997b, Cenarro2003, Schiavon2007, Spiniello2012}. A particular example of such a feature is the CaH1 index defined in \cite{Spiniello2014}. This feature is relatively strong in M-dwarfs but almost absent in M-giants \citep{Spiniello2014}.

Many SPS models fit the spectra of ETGs by assuming one or two single stellar populations (SSPs). In this work we test to what extent this assumption is justified by applying an improved version of the model by DTK16 to a set of realistic composite stellar population (CSP) mock spectra. We use the Bayesian evidence to test how many SSPs are typically necessary to describe a given CSP. Assuming an IMF that varies with velocity dispersion according to \cite{Spiniello2014}, we investigate if we can recover this velocity dispersion--IMF relation from our mock spectra. The outline is as follows. In Section 2 we describe how we extend the DTK16 SSP model to a fully Bayesian CSP model, which uses the evidence for model comparison. In Section 3 we describe the construction of our stellar templates. We specify the CSP mock spectra in Section 4. In Section 5 we report our results and conclusions.

\section{Model description}
\label{sec:modeldescription}
In this work we use the hierarchical Bayesian framework developed by DTK16 for reconstructing the IMF of SSPs. Here, we extend that work with a number of new features that allow us to model the IMF of CSPs. 

\subsection{Hierarchical Bayesian framework}
\label{sec:BayesianFramework}
The basic assumption of the hierarchical Bayesian framework developed by DTK16 is that the spectrum of a galaxy can be written as the sum of the spectra of all the stars that it contains. This is expressed by the following matrix equation:
\begin{equation}
\label{eq:linAssumption}
\mathbfit{g} = \mathbfss{S}\, \mathbfit{w},
\end{equation}
in which $\mathbfit{g}$ is a column vector containing the intensity as a function of wavelength for a population of stars (e.g. a galaxy), $\mathbfss{S}$ is a matrix where each of the columns is formed by the spectrum of a stellar template in the same format as $\mathbfit{g}$, and $\mathbfit{w}$ is a column vector with the number of stars that we have for each of the template stars. We do not account for dust obscuration nor emission lines.

The stars that are present in an SSP are defined by an isochrone as a function of its age and metallicity.  When we combine a stellar library with an interpolator that allows us to interpolate between the spectra of the stars in the library (as a function of effective temperature, surface gravity and metallicity), we can create a spectrum for each of the isochrone stars. For an SSP, these are the spectra that form the columns of matrix $\mathbfss{S}$. We normalize the spectra in matrix $\mathbfss{S}$ to match the Johnson V-magnitude defined by the isochrone. Because an isochrone is a function of age $t$ and metallicity $\mathrm{[M/H]}$, the matrix is of the form $\mathbfss{S}(t,\mathrm{[M/H]})$. The abundance pattern is currently not a free parameter in our model \citep[see][for a model where the abundance of various elements is allowed to vary]{Conroy2012a}.

The vector $\mathbfit{w}$ (referred to as weights) are the unknowns in Equation \ref{eq:linAssumption}. The IMF,
\begin{equation}
\xi (M) \equiv \frac{\mathrm{d}N}{\mathrm{d}M},
\end{equation}
and the weights are related through
\begin{equation}
\label{eq:IMFweights1}
\xi (m_j) = \frac{\mathbfit{w}_j}{m_{\mathrm{high}} - m_{\mathrm{low}}},
\end{equation}
and vice versa through
\begin{equation}
\label{eq:IMFweights2}
\mathbfit{w}_{j} = \int\limits_{m_{\mathrm{low}}}^{m_{\mathrm{high}}} \xi(M) \mathrm{d}M.
\end{equation}
In these equations, $\mathbfit{w}_j$ is the number of stars for template $j$, $m_j$ is the mass of that template star and $m_{\mathrm{low}}$ and $m_{\mathrm{high}}$ are the mass boundaries of this template.

Since Equation \ref{eq:linAssumption} is in general an ill-posed problem, solving it directly through a linear inversion may lead to very unrealistic and unphysical solutions. DTK16 therefore use a prior on the IMF to regulate the inversion of Equation \ref{eq:linAssumption}. Using Equation \ref{eq:IMFweights2}, this IMF prior can be translated into a prior on the weights $\mathbfit{w}_0$. Assuming Gaussian distribution functions of both the errors on the data and the variation of the weights around the mean of the prior, we can solve Equation \ref{eq:linAssumption} by minimizing
\begin{equation}
\begin{aligned}
\label{eq:eqED}
\begin{split}
M(\mathbfit{w}|\mathbfit{g}, \mathbfss{S}, \mathbfit{w}_0, \mathbfss{C}^{-1}_{\mathrm{pr}}) & = \frac{1}{2}(\mathbfss{S}\, \mathbfit{w} - \mathbfit{g})^{\mathrm{T}} \mathbfss{C}_{\mathrm{D}}^{-1} (\mathbfss{S}\, \mathbfit{w}-\mathbfit{g}) \\
& + \lambda \cdot \frac{1}{2} (\mathbfit{w}- \mathbfit{w}_0)^{\mathrm{T}} \mathbfss{C}^{-1}_{\rm{pr}} (\mathbfit{w} - \mathbfit{w}_0),
\end{split}
\end{aligned}
\end{equation}
where $\mathbfss{C}_{\mathrm{D}}^{-1}$ is the inverse noise covariance matrix and $\lambda$ is the regularization parameter that balances the solution of $\mathbfit{w}$ between that of the most likely solution and that of the prior solution. $\mathbfss{C}^{-1}_{\rm{pr}}$ is the inverse covariance matrix describing the shape of the Gaussian distributed deviations of the weights from the prior on the weights $\mathbfit{w}_0$ and is related to the regularization scheme that is used. In this equation, the first term is related to the likelihood and the second (regularization) term puts a penalty on $\mathbfit{w}$ for deviating from the prior on the weights $\mathbfit{w}_0$. The balance between these two terms is set by the regularization parameter $\lambda$ which is chosen such that it optimizes the posterior probability. 

As shown by DTK16, the IMF prior $\mathbfit{w}_0$ can be chosen as part of an IMF prior model family that is defined by a set of non-linear parameters. For example, we may assume that the IMF prior is a power law IMF which is defined by two non-linear parameters: its slope and its normalization. These non-linear IMF prior parameters can be sampled via Markov Chain Monte Carlo (MCMC) sampling techniques, and different choices of IMF prior parameters may be compared on the basis of the Bayesian evidence \citep{MacKay1992}.

At the highest level of inference (second level of inference), we can marginalize over the weights and the non-linear parameters age, metallicity and IMF prior parameters to calculate the evidence for one particular model choice. This allows us to objectively compare different model ingredients, such as the choice of isochrones or the IMF prior model family, with each other on the basis of the evidence.

For more details with respect to the hierarchical Bayesian framework for SPS we refer the reader to DTK16.

\subsection{New model features}
In this section we describe the new features that we add to the model of DTK16. Among these features are a different method for enforcing positive weights, a multiplicative polynomial, an adaptive covariance matrix, the inclusion of multiple SSPs in our data model and including the velocity dispersion as a free parameter of our model. We also discuss a new sampling strategy where we first use a parameterized version of our code to find the most probable ages and metallicities of the SSPs included in the fit. Then we use the full version of the code to reconstruct the parameters related to the IMF.

\subsubsection{Enforcing positive weights}
A negative number of stars is unphysical, and therefore we force the model to consider solutions with positive weights only. DTK16 enforce positive weights by using a non-negative least squares (NNLS) method. We have found that using NNLS can sometimes result in unrealistic solutions. We therefore choose to use a different approach for enforcing positive weights.

If the regularization parameter in our model is increased, we force the solution to become more and more equal to the prior on the weights (which is always positive). Therefore, above some threshold value of the regularization parameter the weights will always be positive. Here, we use this property of the regularization parameter and enforce positive weights by using an additional prior on the regularization parameter $\lambda$. This prior has a probability of one if all weights are positive and zero probability if any of the weights is below zero. With the additional prior on the regularization parameter we ensure a positive solution that is relatively close to the prior IMF while the model still has the freedom to deviate from the IMF prior if required by the data.

\subsubsection{Multiplicative polynomial}
Extinction, flux-calibration issues and systematic uncertainties in the model can potentially introduce differences between the continuum of the input spectrum and the continuum of the model spectrum. A possible solution to absorb these differences is to include a multiplicative polynomial in the fit. The relevance of including a multiplicative polynomial in the fit is discussed in e.g. \cite{Koleva2008}.

Within the context of the hierarchical Bayesian framework discussed in Section \ref{sec:BayesianFramework}, the inclusion of  a multiplicative polynomial is complicated by the fact that the linear weights and the coefficients of the polynomial are degenerate. The weights and polynomial coefficients can therefore not be solved independently. To resolve this problem, we use an iterative procedure.

First, we determine the most probable weights without a multiplicative polynomial. These most probable weights are used to create a reconstructed spectrum and the fractional residual between the reconstructed spectrum and the input spectrum is determined. Then, we fit a multiplicative polynomial to the fractional residual. We apply this polynomial to the spectra in matrix $\mathbfss{S}$ and determine the most probable weights again, now including the effect of a multiplicative polynomial.

To fit the residuals between the reconstructed spectrum and the input spectrum we use a tenth order (Legendre) polynomial $P_{10}(\lambda)$. According to \cite{Koleva2009}, the optimal order of the polynomial depends on wavelength range and resolution but in most cases $n = 10$ should be sufficient to get an unbiased determination of the parameters of a non-linear component, such as for example a stellar atmosphere or an SSP. In addition, \cite{Koleva2009} did not find a degeneracy between their multiplicative polynomials and the parameters of their non-linear components. We have also tested this for a number of SSPs that we distorted by a random polynomial. The results of this test suggest that the inference of age, metallicity and IMF slope of these SSPs is not affected by this distortion.

\subsubsection{Age-metallicity reconstruction}
\label{sec:agemetallicityRec}
To determine the most probable age and metallicity of an SSP, DTK16 consider a grid of ages and metallicities. For each of the points in this grid, the Bayesian evidence is determined by marginalizing over the IMF prior parameters. Under the assumption of a clear peak in the evidence of this age-metallicity grid and only minor degeneracy between the age, metallicity and IMF over the bulk of the evidence in age-metallicity space, the posterior of the weights and IMF prior parameters for the most probable age-metallicity grid point is used as a good approximation for the posterior of the weights and IMF prior parameters marginalized over all ages and metallicities.

Here we use a more general approach where we sample the age and metallicity of an SSP together with the IMF prior parameters. In principle, for every age-metallicity combination we have to create a set of stellar templates. However, since this process is computationally expensive, we use a discrete grid of ages and metallicities for which we a priori create a corresponding set of stellar templates. Within the model, any given combination of age and metallicity is then mapped to the closest discrete age-metallicity grid point and the corresponding set of stellar templates is used for the reconstruction. So although age and metallicity are still internally handled as discrete parameters within the model, this allows us to sample them as continuous parameters. The spacings of the age-metallicity grid points determine our resolution along these dimensions.

\subsubsection{Adaptive covariance matrix}
\label{sec:extraCovariance}
The standard noise covariance matrix in our model only takes into account the observational error on the input spectrum. In reality, there are additional uncertainties such as systematics in the model and uncertainties related to the input spectrum being a CSP instead of an SSP. Therefore, we will in general underestimate the error budget of the model parameters. To model these additional uncertainties, we introduce an additional variance with value $b$. This parameter represents additional uniform variance that we do not take into account. If the original (diagonal) covariance matrix is given by $\mathbfss{C}_{\mathrm{D, old}}$, the new covariance matrix becomes
\begin{equation}
\mathbfss{C}_{\mathrm{D,new}} = \mathbfss{C}_{\mathrm{D,old}} + b \mathbfss{I}.
\end{equation}

Note that this is still a diagonal covariance matrix. When we apply our model to real data, a more general covariance matrix, such as for example discussed by \cite{Czekala2015}, might be required. However, since we need to calculate the inverse of the covariance matrix, which can be rather time consuming, we do not currently implement such a general covariance matrix.

\subsubsection{Multiple SSPs}
Real galaxies are CSPs. If we model a galaxy as if it is an SSP, this is an approximation that can introduce additional uncertainties. Depending on the star formation history (SFH) of the galaxy that we consider, this may or may not be a good approximation. The  ETGs that we are interested in are in general expected to be characterized by an SFH consisting of an initial peak (old stellar population) followed by a small amount of more recent star formation \citep[as observed in present-day ETGs by, e.g.][]{Kaviraj2008, Monachesi2012, McDermid2015}. Although the initial peak is probably modelled reasonably well by an SSP, the SFH is in reality extended. This may introduce a bias on the inferred age and metallicity of the SSP and potentially affect our inference of the IMF. To absorb some of these effects, we extend our model in such a way that it allows us to model a CSP as a combination of a given number of SSPs with varying age and metallicity. This approach allows us in principle to recover the age-metallicity distribution function that has the highest overall Bayesian evidence. It also allows for objective comparison between SSP models and CSP models with a varying number of SSP components (see also Section \ref{sec:reqNssps} where we test this for a number of mock CSP spectra).

Compared to our original approach, where the columns of matrix $\mathbfss{S}$ correspond to the spectra of an SSP, the columns of matrix $\mathbfss{S}$ now correspond to the spectra of multiple SSPs. For example, for $N$ SSPs matrix $\mathbfss{S}$ becomes
\begin{equation}
\mathbfss{S} = \left(\,
\begin{array}{  c c c c c }
{} & {} & {} \\
\mathrm{SSP}_1 & \cdots & \mathrm{SSP}_i & \cdots & \mathrm{SSP}_N\\
{} & {} & {} \\
\end{array}\,\right).
\end{equation}
Similarly, the weights in Equation \ref{eq:linAssumption} become
\begin{equation}
\mathbfit{w} = \left(\,
\begin{array}{c}
\mathbfit{w}_{\mathrm{SSP}_1} \\
\vdots \\
\mathbfit{w}_{\mathrm{SSP}_i} \\
\vdots \\
\mathbfit{w}_{\mathrm{SSP}_N} \\
\end{array}\,\right),
\end{equation}
and the prior on the weights becomes
\begin{equation}
\mathbfit{w}_0 = \left(\,
\begin{array}{c}
\mathbfit{w}_{0,\mathrm{SSP}_1} \\
\vdots \\
\mathbfit{w}_{0,\mathrm{SSP}_i} \\
\vdots \\
\mathbfit{w}_{0,\mathrm{SSP}_N} \\
\end{array}\,\right).
\end{equation}

In principal, for every SSP that we take into account we can have a different IMF prior $\mathbfit{w}_{0, \mathrm{SSP}_i}$. Here, we assume that the shape of the IMF prior is the same for all SSPs that we include in our model. However, the normalization of the IMF prior is a free parameter for every SSP because every SSP will contribute differently to the integrated light of the CSP. Note that although the shape of the IMF prior is assumed to be the same for all SSPs, the actual inferred IMF shape can change for different SSPs since the most probable weights are allowed to deviate from the prior for each SSP.

Altogether, every SSP introduces three additional non-linear parameters that have to be sampled by the MCMC sampler: its age, its metallicity and the normalization of the IMF prior, besides the IMF shape parameters which are assumed to be the same for all populations (hence no additional free parameters are introduced as compared to an SSP). The computational time scales are related to the increase in the number of parameters that have to be sampled and the increase of matrix $\mathbfss{S}$ as we add more SSPs, which limits this approach to a few SSPs.

\subsubsection{Velocity dispersion}
In DTK16, we smooth our (mock) spectra to a given velocity dispersion. The stellar templates are then smoothed to the same velocity dispersion. Here, we include the velocity dispersion $\sigma$ as an additional free parameter. We assume a Gaussian smoothing kernel in $\log \lambda$.

\subsubsection{Parameterized version of model}
\label{sec:likVersion}
In the original version of the model, we use the prior on the weights $\mathbfit{w}_0$ to regularize the linear inversion of the most probable weights $\mathbfit{w}_{\mathrm{MP}}$. However, for a given set of stellar templates $\mathbf{S}$ we can also use $\mathbfit{w}_0$ to directly compute a corresponding prior spectrum $\mathbfit{g}_0$ through
\begin{equation}
\label{eq:priorSpec}
\mathbfit{g}_0 = \mathbfss{S}\, \mathbfit{w}_0.
\end{equation}
This prior spectrum can be compared to the data $\mathbfit{g}$ on the basis of the likelihood
\begin{equation}
\label{eq:likelihood}
{\cal{L}} (\mathbfit{g}_0|\mathbfit{g}, \mathbfss{S}) = \frac{\mathrm{exp}[-E_{\!_D}(\mathbfit{g}_0|\mathbfit{g}, \mathbfss{S})]}{Z_{\!_D}},
\end{equation}
in which $Z_{\!_D}$ is the normalization of the likelihood and 
\begin{equation}
E_{\!_D}(\mathbfit{g}_0|\mathbfit{g}, \mathbfss{S}) = \frac{1}{2}(\mathbfit{g}_0 - \mathbfit{g})^{\mathrm{T}} \mathbfss{C}_{\mathrm{D}}^{-1} (\mathbfit{g}_0-\mathbfit{g}).
\end{equation}

We have developed a version of the code where we compare the input spectrum $\mathbfit{g}$ directly to the prior spectrum $\mathbfit{g}_0$ on the basis of the likelihood in Equation \ref{eq:likelihood}. We will refer to this version of the code as the parameterized version. The advantage of the parameterized version of the the code is that it is much faster then the original version. In the original version of the code, every iteration requires an iteration to find the most probable regularization parameter, and, for every step in this iteration, we have to perform a linear inversion to determine the most probable weights and the evidence. This process is time consuming. In the parameterized version of the code we only evaluate Equations \ref{eq:priorSpec} and \ref{eq:likelihood}, which is much faster but comes at the cost of less flexibility.

Similar to the full version of the code, we include a multiplicative tenth-order (Legendre) polynomial in the fit. In Section \ref{sec:samplingStrategy} we explain when we use the parameterized version of the code and when we use the full version of the code in our sampling strategy.

\subsubsection{Sampling strategy}
\label{sec:samplingStrategy}
In the CSP version of the code we change our sampling strategy. First, we use the parameterized version of our code discussed in Section \ref{sec:likVersion} to determine the most probable ages and metallicities of the SSPs that we include in our fit. As discussed in Section \ref{sec:agemetallicityRec}, the ages and metallicities of the SSPs are sampled together with the IMF prior parameters. Then, we fix the ages and metallicities of the SSPs to the most probable values from the first step and we sample the IMF prior parameters with the full version of our model (including linear inversion). This initial sampling procedure is much faster than when we sample the complete space of ages, metallicities and IMF prior parameters with the full version of our model but assumes that there is little covariance between the age-metallicity parameters and small deviations from the IMF prior. We also use the parameterized version of our code to determine the velocity dispersion of a spectrum.

As in DTK16, we have implemented our code as a pipeline of \texttt{cosmoSIS} \citep{Zuntz2015}. We use \texttt{Multinest} \citep{Multinest2008, Multinest2009, Multinest2013} to calculate evidences and obtain a posterior sample of the sampled parameters. If necessary, the posterior sampling is refined with \texttt{emcee} \citep{EMCEE}.

\section{Stellar templates}
In this section we describe how we construct the stellar templates that form the columns of the stellar-template matrix $\mathbfss{S}$. We also describe a binning procedure where we combine multiple stellar templates into one new template. This binning procedure allows us to reduce the sizes of the stellar templates matrices and to speed up the code.

\subsection{Isochrones}
The stellar templates that we use in this work are created on the basis of the stellar parameters provided by the Parsec isochrones \citep{Bressan2012} extended with the thermally pulsating - asymptotic giant branch (TP-AGB) tracks from \cite{Marigo2013} and \cite{Rosenfield2016}. This is the most recent version of the Parsec isochrones and includes several updates to the input physics of the Padova stellar evolution code, as well as an improved treatment of low mass dwarfs as described in \cite{Chen2014}. Each isochrone provides us with the initial masses, effective temperatures, surface gravities and Johnson V-band magnitudes of the stars that are present in an SSP.

\subsection{Stellar library and interpolator}
To transform the stellar parameters of an isochrone into a set of spectra, we combine a stellar library with an interpolator. We use the empirical MILES stellar library \citep{MILES} which consists of approximately 1000 stars in the wavelength range 3540.5 - 7409.6 \AA. To interpolate between the stars in this library, we use the same interpolator as in DTK16. This is a local interpolator based on \cite{Vazdekis_interpolator}. Since we now include a multiplicative polynomial in our model, we do not apply the polynomial correction of the MILES stars discussed in DTK16 but use the spectra of the MILES library directly.

The interpolator $S_{\lambda}(T_{\mathrm{eff}}, \log g, [\mathrm{Fe/H}])$ allows us to create a stellar spectrum for any given set of stellar parameters $T_{\mathrm{eff}}$, $\log g$ and $\mathrm{[Fe/H]}$. With this interpolator we create a stellar-template matrix $\mathbfss{S}$ for each of the grid points in our age-metallicity grid. Since the isochrones are defined on the basis of $[\mathrm{M/H}] = \log(Z/Z_{\odot})$ but the MILES stars only have measured values of $[\mathrm{Fe/H}]$, we assume that $[\mathrm{M/H}] = [\mathrm{Fe/H}]$.

\subsection{Isochrone binning}
\label{sec:isochroneBinning}
Since the parameters of one isochrone star to the next isochrone star do in general not change drastically, the interpolated spectra are expected to vary smoothly as a function of stellar mass. The number of stars in a Parsec isochrone is typically on the order of a few hundred and can be as high as a thousand. Having such a large number of templates makes the matrices in our model unnecessarily large which in turn makes the code unnecessary slow. We therefore reduce the number of stellar templates in the matrices by binning them together. 

The sampling of the isochrone stars increases as a function of evolutionary phase and is particularly high for the TP-AGB. Hence, we let the number of stars that we bin together depend on the evolutionary phase. For the main sequence we combine every two spectra into a new template, for the TP-AGB we combine every eight spectra into a new template and for the other evolutionary phases we combine every three spectra into a new template. With these numbers, we reduce the computational time scales significantly and are still able to model a spectrum with up to $N=6$ SSPs (see Section \ref{sec:reqNssps}) in a reasonable amount of time\footnote{Typically, modelling a spectrum with $N = 6$ SSPs takes on the order of 3-4 hours using 64 cores.}.

When we bin the spectra, we assume a Salpeter IMF. In Appendix \ref{sec:AppIsochroneBinning} we show that this assumption has only a very small effect on our SSP spectra and that it does not affect our inference of the IMF.

\section{Composite stellar population mock spectra}
\label{sec:spectraConstruction}
In this section we describe the CSP mock spectra that we use to test our model. First, we discuss the semi-analytic model from which we extract the SFHs of our CSP mock spectra. Then we describe how we use these SFHs to construct a set of CSP mock spectra with a variable IMF.

\subsection{SFHs from semi-analytic models}
Semi-analytic models (SAMs) are an ideal tool to create realistic SFHs of galaxies. The main advantage of SFHs from SAMs over more idealized SFHs (e.g., delayed $\tau$ models) is that they can include increasing and decreasing phases, as well as bursts of star formation. Within SAMs, simplified analytical but physically motivated recipes are used to track dynamical and astrophysical processes. This happens on entire galaxy scales, rather than kpc or even pc scales (the physical scales below which cosmological and zoom-in hydro-simulations apply subgrid physics).  This ensures an inexpensive runtime for SAMs, which makes them a powerful tool to generate statistically significant model galaxy populations covering a wide range in galaxy properties \citep[for a recent review see][]{SomervilleDave2015}.

In this work we focus on the SAM first presented in \citet{Somerville1999} and further updated by \citet{Somerville2008}, \citet{Somerville2012}, and \citet{Porter2014}. This SAM includes recipes to track the hierarchical clustering of dark-matter haloes, shock heating and radiative cooling of gas, supernova feedback, star formation, black-hole growth and active galactic nuclei feedback, the enrichment of the interstellar and intracluster medium with metals, mergers of galaxies, starbursts, and the evolution of stellar populations. \citet{Porter2014} included recipes to track the effect of disk-instabilities and mergers on the structural evolution of galaxies. Of particular importance within the context of this work, \citet{Porter2014} included recipes to track the line-of-sight velocity dispersion of the stars in modelled galaxies. These models have proven to be fairly successful in reproducing local and high-redshift statistical galaxy properties such as stellar mass functions, gas masses, dust masses, sizes, SFRs, and also quenched fractions and structural properties such as S\'ersic index\citep{Somerville2008, Somerville2012, Somerville2015, Porter2012, Popping2014, PoppingSomervilleGalametz2016, PoppingvanKampen2016, Lang2014, Brennan2015, Brennan2017}. We refer the reader to the aforementioned works for more detailed descriptions of the utilized model.

We make use of merger trees extracted from the Bolshoi N-body dark matter simulation \citep{Klypin2011,Trujillo-Gomez2011,Rodriguez-Puebla2016}. Dark matter haloes were identified using the \texttt{ROCKSTAR} algorithm \citep{BehrooziRockstar2013}. This simulation is complete down to haloes with a virial velocity of $V_{\rm circ}=50\,\rm{km}\,\rm{s}^{-1}$, with a force resolution of 1 $h^{-1}$ kpc and a mass resolution of $1.9 \times 10^8\,\rm{M}_\odot$ per particle. The results presented assume a $\Lambda$CDM cosmology with $\Omega_m=0.27$, $\Omega_\Lambda = 0.73$, $h = 0.7$, $\sigma_8=0.82$, $n=0.95$ and a cosmic baryon fraction of $f_b=0.1658$. These parameters were chosen to match those adopted in the Bolshoi simulation \citep{Klypin2011} and are consistent with the \textit{Wilkinson Microwave Anisotropy Probe} (WMAP) results \citep{Komatsu2011}. 

The semi-analytic model assumes a \citet{Chabrier2003} IMF. Note that this IMF is only assumed within the SAM. As described in Section \ref{sec:CSPspectra}, the spectra of the mock CSPs are determined under the assumption of variable IMF. Therefore, the IMF assumed in the SAM and the IMF used to determine the spectra are not consistent with each other. In principal, the assumption of a universal \citet{Chabrier2003} IMF within the SAM might affect the enrichment of the ISM with metals, the mass-loading factor and the heating rate from massive stars (supernovae and winds), and eventually also the stellar mass assembly of a galaxy. However, within the context of this work we only want to construct a complex age-metallicity distribution and not necessarily a precise one and therefore these effects are not expected to affect our conclusions.

Galaxies drawn from the SAM are selected to be passive, with specific star formation rates $\mathrm{SFR}/M_\ast<10^{-11}\,\mathrm{yr^{-1}}$, to match approximately the selection criteria of \citet{Spiniello2014}.  We then average the SFHs of these passive galaxies in $40\,\mathrm{km \,s^{-1}}$-wide bins centred at velocity dispersions of $150$, $190$, $230$, $270$, and $310\,\mathrm{km\,s^{-1}}$. The resulting SFHs suggest that some of the stars that have formed are relatively young ($t < 8$ Gyr) with a very low metallicity ($\mathrm{[Fe/H]} = -2.4$). We suspect that this is connected to a mode of star formation which is related to infalling gas from the intergalactic medium or it could be accretion from a poorly resolved, low-mass satellite. Since this kind of stellar population is not seen in present-day galaxies, we manually set the stellar mass in bins with $t < 8$ Gyr and $\mathrm{[Fe/H] = -2.4}$ to zero.

\cite{TragerSomerville2009} compare SSP-equivalent ages between a mock catalogue of SAM galaxies and a number of observational samples and show that these ages approximately agree. Although the observational samples do not allow the reconstruction of the full SFH, this would imply that the amount of recent star formation produced by the SAM is approximately correct. \cite{Fontanot2009} compare the mean mass-weighted stellar age as a function of stellar mass of three SAMs with the observational results of \cite{Gallazzi2005} and show that the agreement for massive galaxies is very good. \cite{Porter2014b} compare mass weighted ages of ETGs in their SAM with the luminosity-weighted ages of the sample of SDSS ETGs by \cite{Graves2009} as a function of effective radius and velocity dispersion. The SDSS galaxies are found to be younger than the SAM galaxies, but this difference most likely originates from comparing luminosity-weighted with mass-weighted ages, where the former is known to be biased towards younger ages, i.e. dominated by young stars \citep{TragerSomerville2009}. However, the trends recovered between velocity dispersion, age and metallicity are very similar for both samples. These results suggest that the SFHs that we use in this work may not be completely accurate but are at least a realistic representation of what might be expected for ETGs.

\subsection{CSP mock spectra with a variable IMF}
\label{sec:CSPspectra}
\begin{figure*}
	{\includegraphics[width=0.3\textwidth]{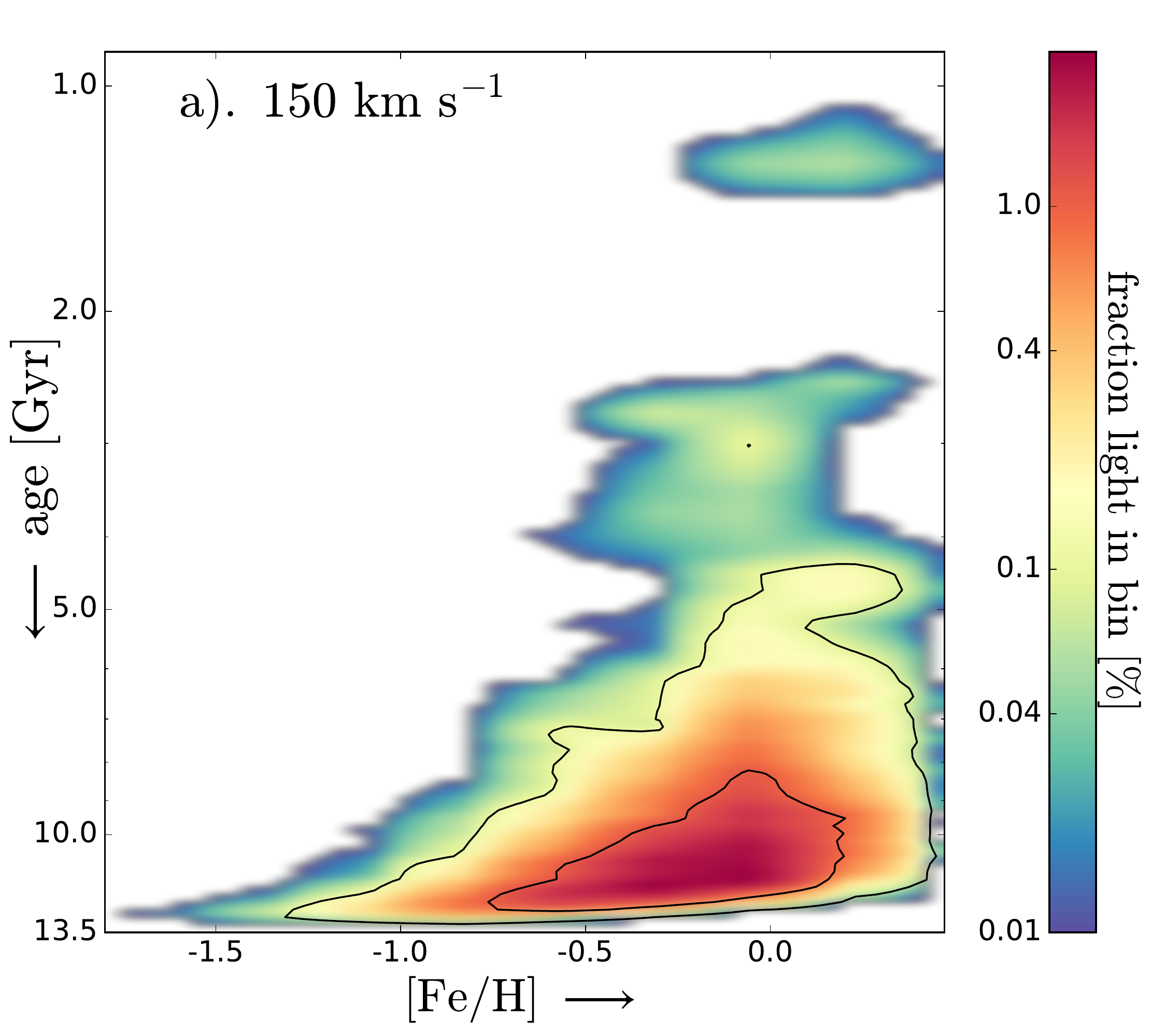}}
	{\includegraphics[width=0.3\textwidth]{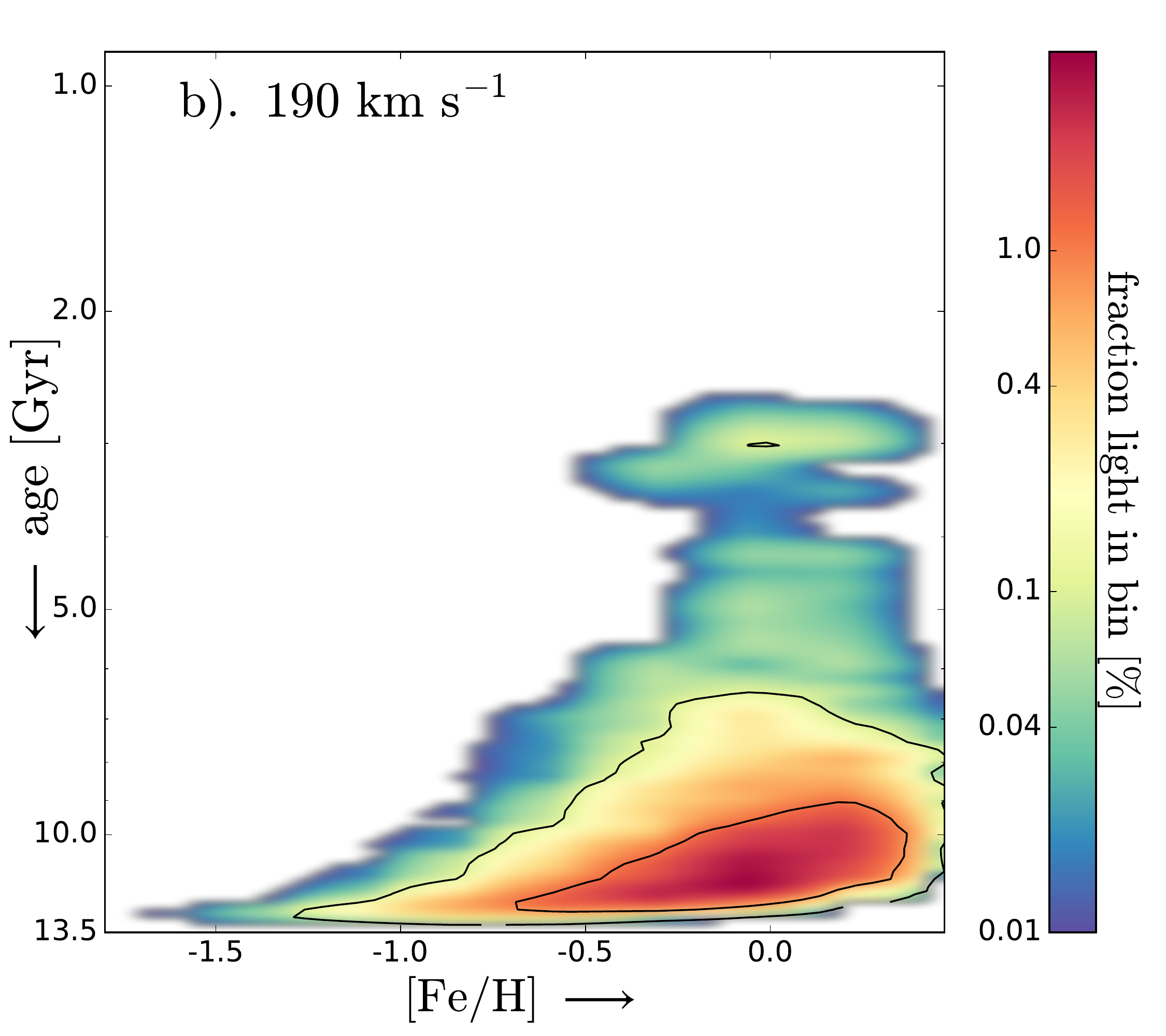}}
	{\includegraphics[width=0.3\textwidth]{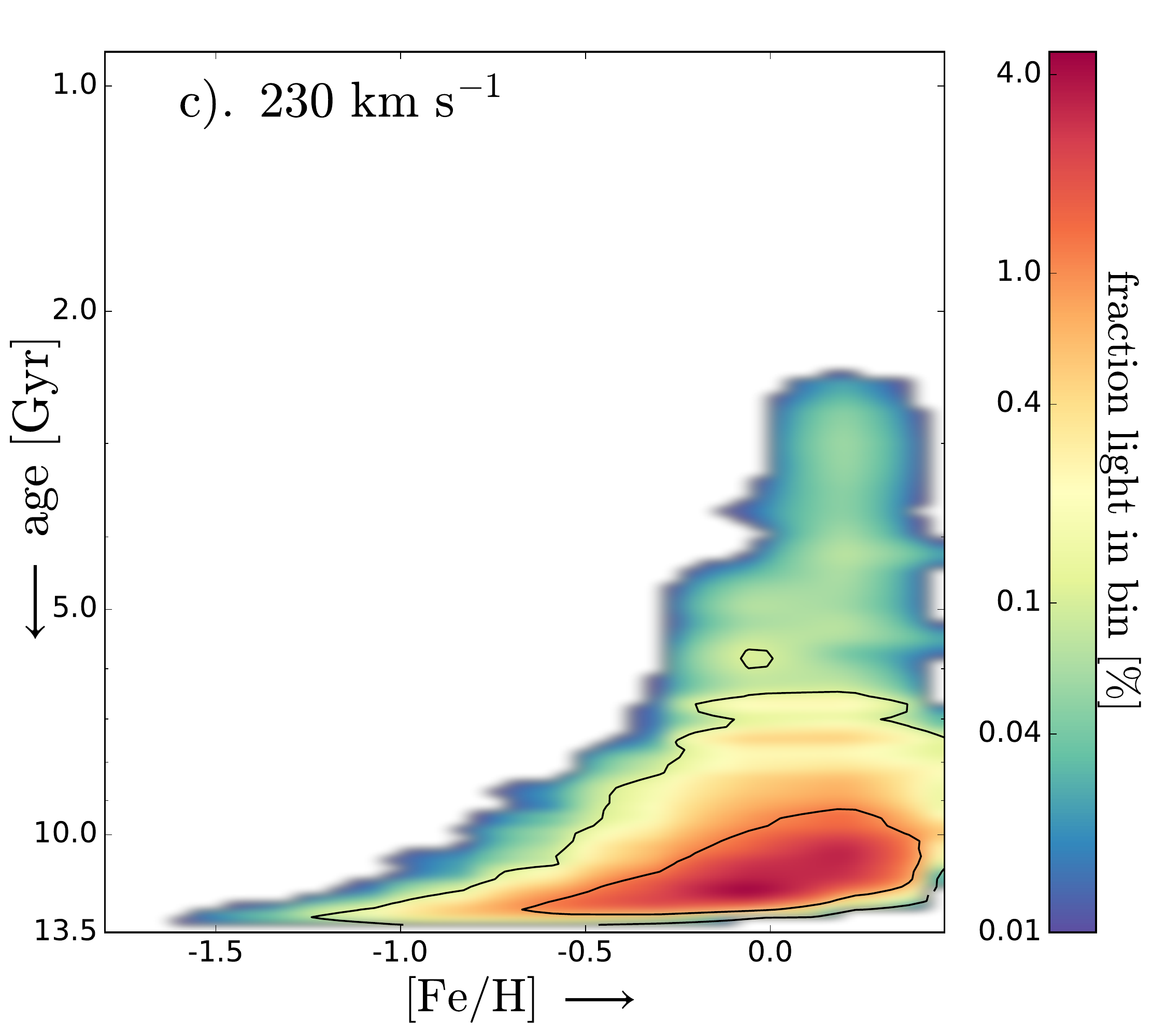}}
	{\includegraphics[width=0.3\textwidth]{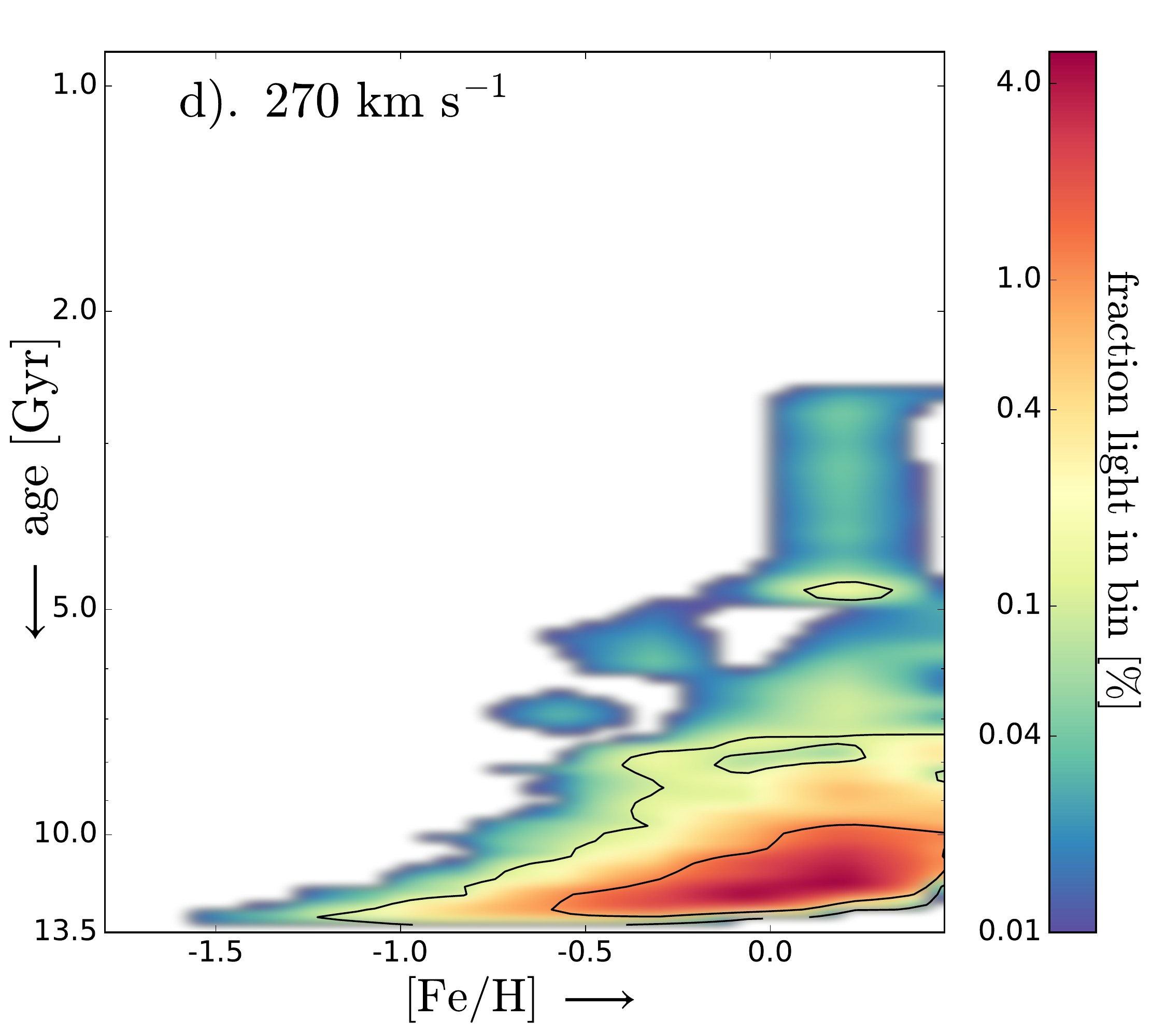}}
	{\includegraphics[width=0.3\textwidth]{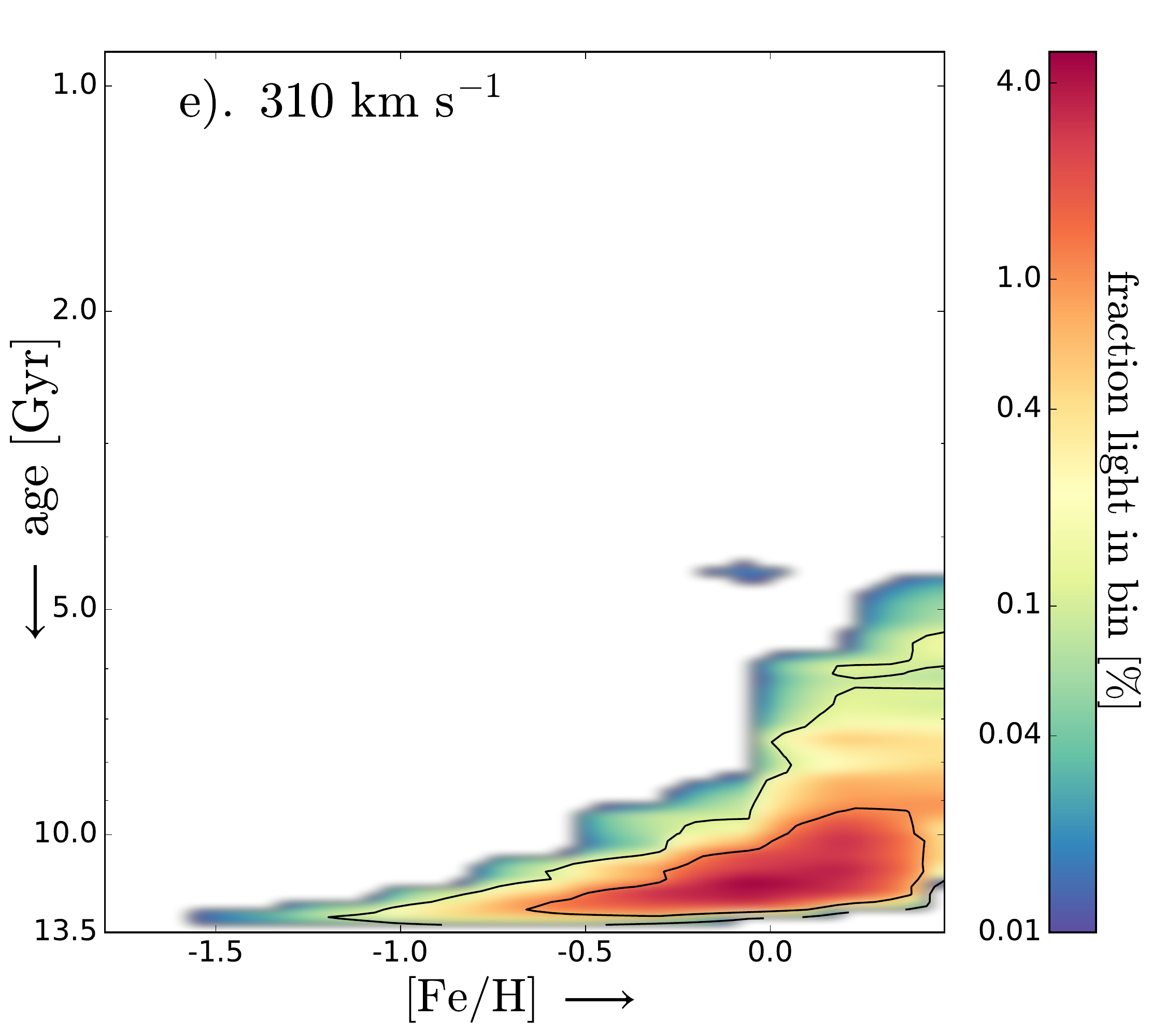}}
	{\includegraphics[width=0.3\textwidth]{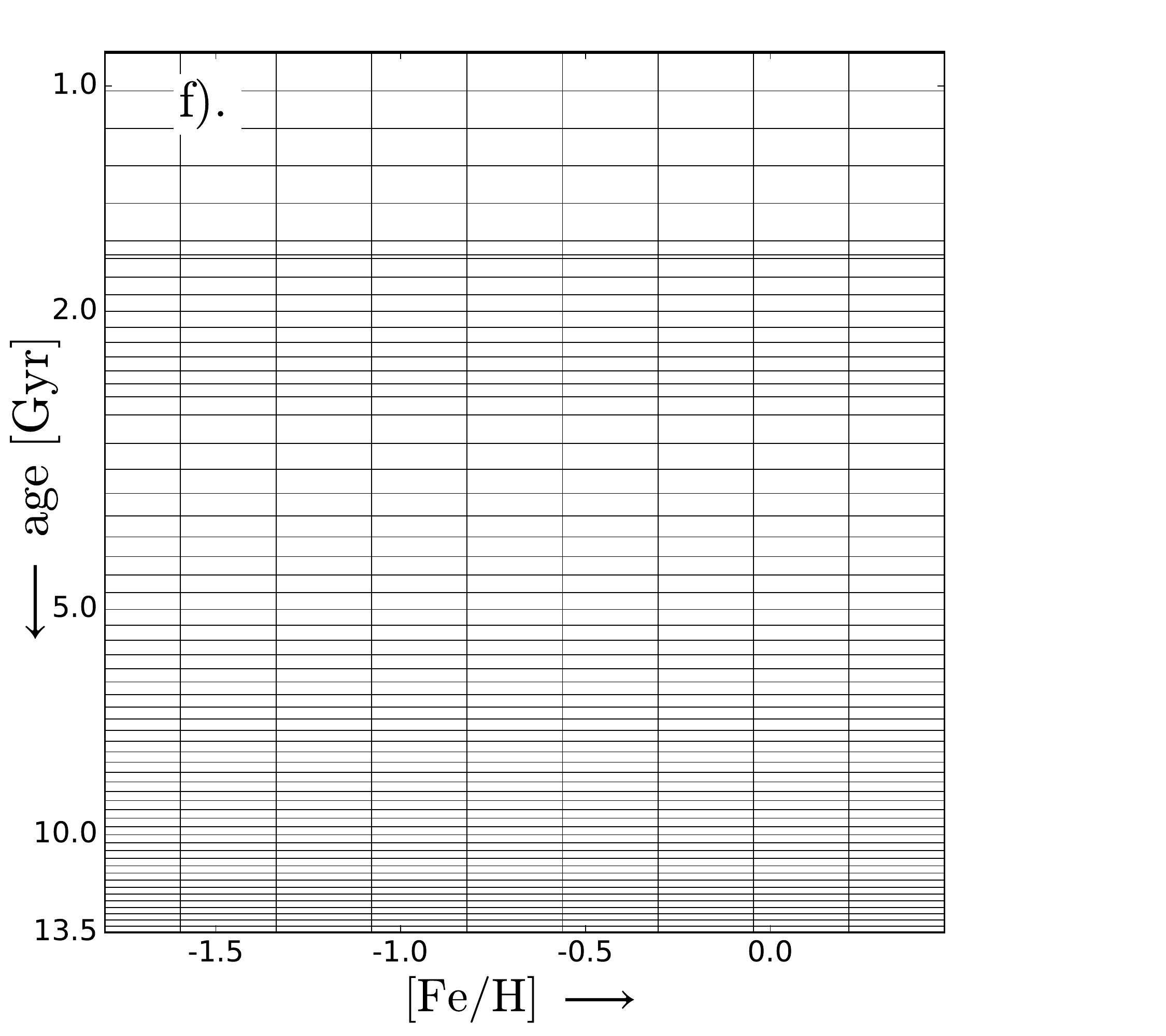}}
	\caption{Average SFHs of galaxies in the SAM binned by velocity dispersion. For each point in the age-metallicity grid, we determine the fractional contribution of the integrated SSP spectrum to the integrated CSP spectrum and the (colour) contours in this plot are interpolated on the basis of these values. The SFHs in panel a-e correspond to different velocity dispersions. Panel (f). represents the adopted age-metallicity grid in the same range as the previous panels.}
	\label{SFHs}
\end{figure*}

The SAMs provide us with an average SFH for galaxies in five different bins of velocity dispersion (150, 190, 230, 270  and $310 \: \mathrm{km} \: \mathrm{s}^{-1}$). The average SFH for the different velocity dispersion bins is visualized in Figure \ref{SFHs}. For each of the velocity dispersion bins, the average SFH is given as the total current mass of stars in a predefined age-metallicity grid. This age-metallicity grid has ages in the range $t = \{0.1, 13.5\}$ Gyr and metallicities in the range $\mathrm{[M/H]} = \log (Z/Z_{\odot}) = \mathrm{[Fe/H]} = \{-2.4,0.47\}$ with $Z_{\odot} = 0.0152$ \citep{Bressan2012}. The spacings of this grid are shown in the last panel of Figure \ref{SFHs}. Note that the age-metallicity grid of the isochrones is chosen to match this grid.

We assume that the IMF of our CSP mock spectra is a single power law IMF. Furthermore, we assume that the slope $\alpha$ of this IMF varies as a function of velocity dispersion according to the empirical IMF slope--velocity dispersion relation of \cite{Spiniello2014}:
\begin{equation}
\label{eq:SpinielloRelation}
\alpha = 2.3 \log \sigma_{200} + 2.13,
\end{equation}
where $\sigma_{200}$ is the velocity dispersion in units of 200 km s$^{-1}$.

For every bin in velocity dispersion we create a corresponding CSP mock spectrum. To create these CSP mock spectra, we first create an SSP spectrum $\mathrm{SSP}_{ij}$ for each of the age-metallicity grid points. These SSP spectra are calculated using Equation \ref{eq:linAssumption}, where the stellar templates $\mathbfss{S} = \mathbfss{S}(t_i, \mathrm{[Fe/H]}_j)$ are those of an SSP and the weights $\mathbfit{w}$ are determined through Equation \ref{eq:IMFweights2} assuming a single power law IMF $\xi(\sigma)$ where the slope is given by Equation \ref{eq:SpinielloRelation}. The SSP spectra are normalized to a total (current) stellar mass of 1 $\mathrm{M}_{\odot}$. 

\begin{table}
	\centering
	\caption{Summary of the five different CSP mock spectra that we consider. Each of the mock spectra corresponds to a different velocity dispersion bin, the SFH is derived from the SAM and the IMF is assumed to be a single power law IMF. The first column gives the name of the CSP mock spectrum, the second column the velocity dispersion bin and the last the column the IMF slope. For every spectrum there are three versions corresponding to SNRs of 75, 150 and 300, respectively.}
	\label{tab:CSPmockspectra}
	\begin{tabular}{ccc} 
		\hline
		name & $\sigma$ [km s$^{-1}$] & $\alpha$\\
		\hline
		\texttt{CSP150} & 150.0 & 1.84\\
		\texttt{CSP190} & 190.0 & 2.08\\
		\texttt{CSP230} & 230.0 & 2.27\\
		\texttt{CSP270} & 270.0 & 2.43\\
		\texttt{CSP310} & 310.0 & 2.57\\	
		\hline
	\end{tabular}
\end{table}

The CSP mock spectrum $\mathbfit{s}_{\mathrm{CSP}}$ is then determined as the weighted sum over all SSP spectra:
\begin{equation}
\mathbfit{s}_{\mathrm{CSP}} = \sum\limits_{i,j} c_{ij}\, \mathrm{SSP}_{ij},
\end{equation}
where the weights $c_{ij}$ of each of the SSP spectra follow from the SFHs of the SAM. We smooth the CSP mock spectra according to the corresponding model velocity dispersion. This gives us a total of five mock CSP spectra, which are summarized in Table \ref{tab:CSPmockspectra}. As a final step, we add Gaussian noise to the spectra to mimic observations with signal-to-noise ratios (SNRs) of 75, 150 and 300 per bin, so that we have three version of every CSP mock spectrum corresponding to different SNRs.

\section{Results}
In this section we present the analysis of the CSP mock spectra with our model. We focus first on \texttt{CSP150} and \texttt{CSP310}. These mock spectra are characterized by, respectively, the most and least extended SFH, as can be seen in Figure \ref{SFHs}. Then we extend our analysis to the other mock spectra and show that we can reconstruct the underlying variable IMF. We use the sampling strategy discussed in Section \ref{sec:samplingStrategy}. First we use the parameterized version of our model to determine the most probable ages and metallicities of the SSPs and the velocity dispersion of the spectrum. Note that the velocity dispersion is derived by broadening the stellar templates in matrix $\mathbfss{S}$ with $\sigma$ as a free parameter. Then we use the full version of the model to sample the non-linear IMF prior parameters and the additional covariance parameterized by $b$ as discussed in Section \ref{sec:extraCovariance}.

For all of the analyses that we present here, we assume a single power law IMF prior. The regularization scheme that we use penalizes the relative deviation of the weights $\mathbfit{w}$ from the prior on the weights $\mathbfit{w}_0$ and is specified by
\begin{equation}
\mathbfss{C}^{-1}_{\mathrm{pr},ii} = \frac{1}{{\mathbfit{w}_{0,i}}^2}
\end{equation}
where $\mathbfss{C}^{-1}_{\mathrm{pr}}$ is a diagonal matrix.

\subsection{The required number of SSPs}
\label{sec:reqNssps}
The Bayesian context of our model allows us to determine the most probable number of SSPs that is required to fit the data. As discussed in Section \ref{sec:modeldescription}, our model is now able to fit a spectrum with a combination of multiple SSPs. For a given number of SSPs, if we marginalize over all the free parameters in the model we can determine the evidence for the given number of SSPs. Running the model multiple times allowing for different numbers of SSPs therefore allows us to evaluate the Bayesian evidence as a function of the number of SSPs included in the fit. In principle, one can do this for an arbitrary number of SSPs. However, due to computational costs we limit ourselves to a maximum of $N = 6$ SSPs.

\begin{figure}
	\centering
		{\includegraphics[width=0.8\columnwidth]{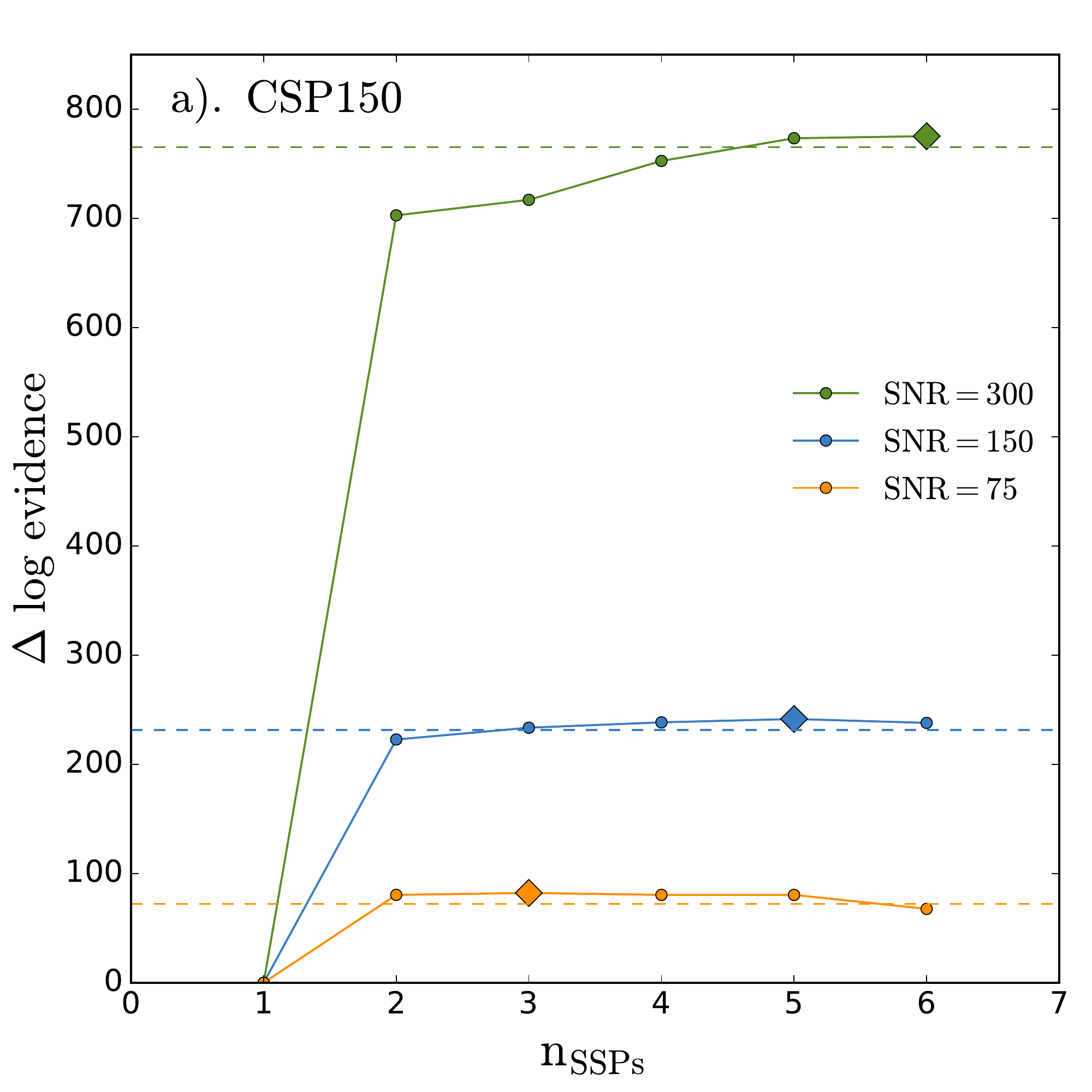}}
		{\includegraphics[width=0.8\columnwidth]{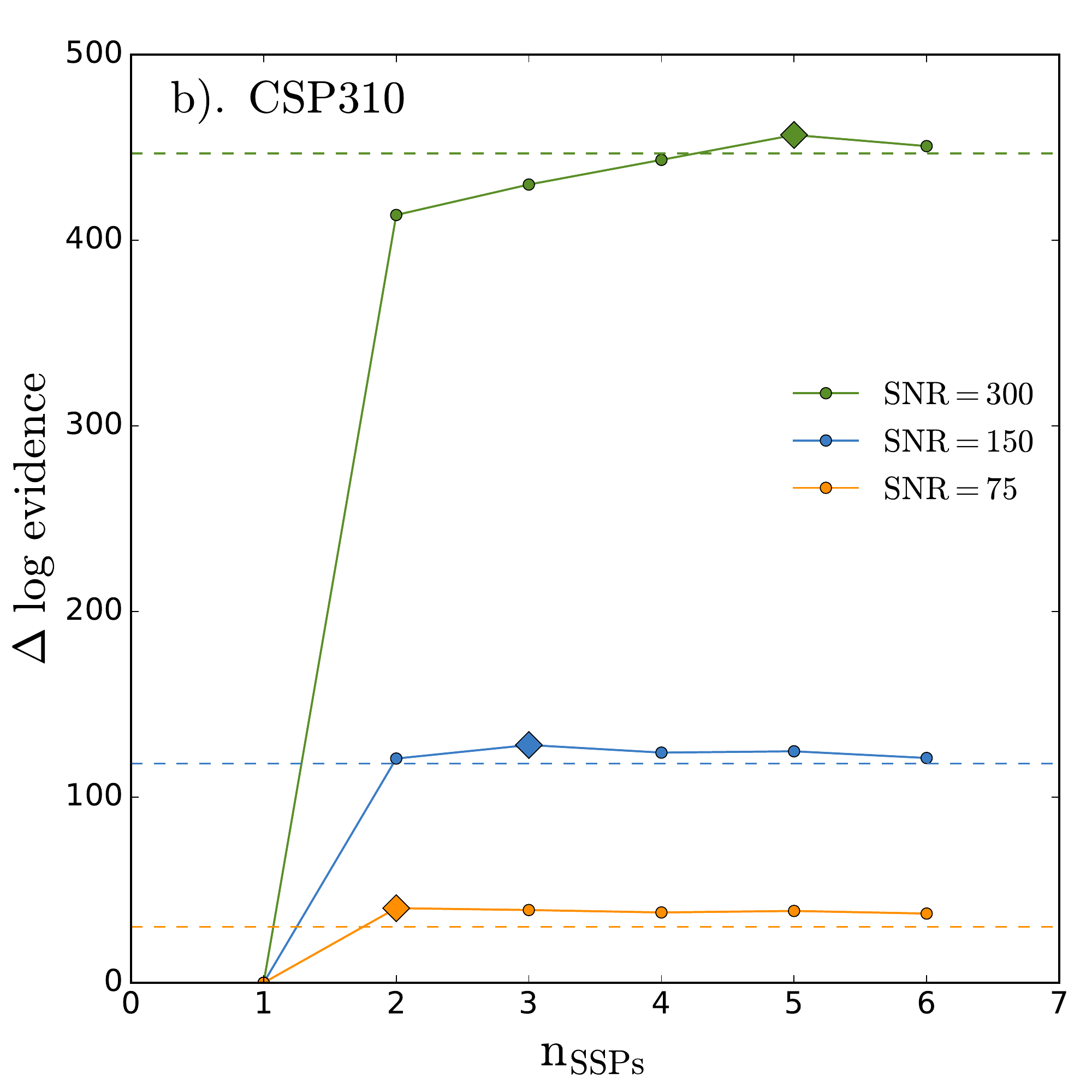}}
	\caption{Difference in log evidence as a function of the number of SSPs included in the analysis. The difference in log evidence is determined with respect to the evidence obtained with one SSP such that for $n_{\mathrm{SSPs}} = 1$, $\Delta \log \mathrm{evidence} = 0$. Panel (a) corresponds to \texttt{CSP150} and panel (b) to \texttt{CSP310}. Different lines correspond to different SNRs as indicated in the legend. The diamond data point represent the number of SSPs for which the evidence reaches a maximum. The horizontal lines correspond to a difference of $\Delta \log \mathrm{evidence} = -10$ with the peak value of the evidence, indicating for how many SSPs there is strong evidence to include them in the fit. For example, there is strong evidence for including five SSPs for \texttt{CSP150} at SNR=300 while at SNR=75 there is only strong evidence for including two SSPs.}
	\label{SSPevidence}
\end{figure}

Figure \ref{SSPevidence} shows the difference in log evidence for \texttt{CSP150} and \texttt{CSP310} as a function of the number of SSPs and compared to the evidence determined with one SSP. The different lines in Figure \ref{SSPevidence} correspond to different SNRs, and one can clearly see that the most probable number of SSPs depends on the SNR of the spectrum. For SNR=75, the most probable number of SSPs is about 2-3, while for SNR=300 the evidence continues to rise to about 5-6 SSPs. This SNR dependence is not surprising, since spectra with higher SNR contain more information and are in general better able to discriminate between models.

In addition to the SNR dependence, Figure \ref{SSPevidence} also shows that the most probable number of SSPs for \texttt{CSP310} is lower than it is for \texttt{CSP150} for spectra with the same SNR. This difference is most likely explained by the more extended SFH of \texttt{CSP150}. In general, we expect that the more complicated the SFH the more SSPs are required to correctly fit the spectrum of a CSP. Another explanation might be the higher velocity dispersion of \texttt{CSP310}. A higher velocity dispersion implies that more information in the spectrum is lost due to the smoothing, and as a consequence it might become more difficult to distinguish different SSPs.

Figure \ref{SSPevidence} shows that we can gain significantly by going from one SSP to two SSPs. For all the CSP mock spectra that we consider, the difference in log evidence between two subsequent number of SSPs is the largest for going from one to two SSPs and ranges from a difference in log evidence of approximately 25 to approximately 700. According to \cite{Jeffreys1961}, a difference in log evidence of more than ten can be considered as strong evidence in favour of a particular model. Therefore, for all SSPs that we considered there is strong evidence to include two SSPs instead of one. Allowing for more than two SSPs may still improve the evidence, but the differences in log evidence are not as conclusive as going from one to two SSPs, though this might strongly depend on the SFH. The horizontal lines in Figure \ref{SSPevidence} indicate a difference of $\Delta \log \mathrm{evidence} = -10$ with the peak value of the evidence. After the evidence curve crosses this horizontal line for the first time, there is no strong evidence to include any additional SSPs in the fit. Typically, for SNR = 75 there is strong evidence to include 2 SSPs, for SNR = 150 there is strong evidence to include 2-3 SSPs and for SNR=300 there is strong evidence to include 5 SSPs. (See also Appendix \ref{app:systematicUncertainties} for a test where we reconstruct the spectra of three mock SSPs with and without the original stellar templates to assess the level of the effect of systematic uncertainties on the difference in log evidence.)

\begin{figure*}
	{\includegraphics[width=0.3\textwidth]{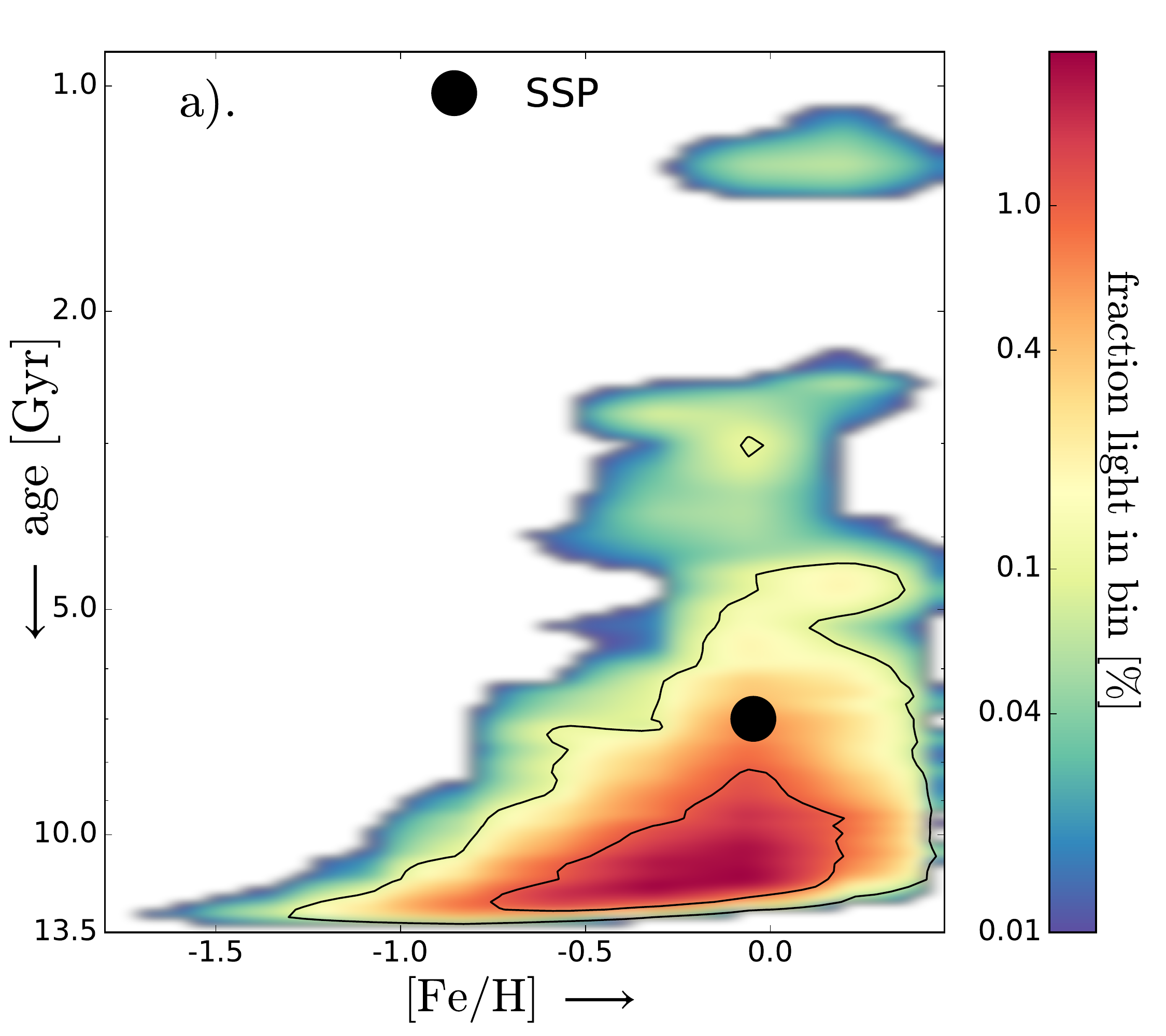}}
	{\includegraphics[width=0.3\textwidth]{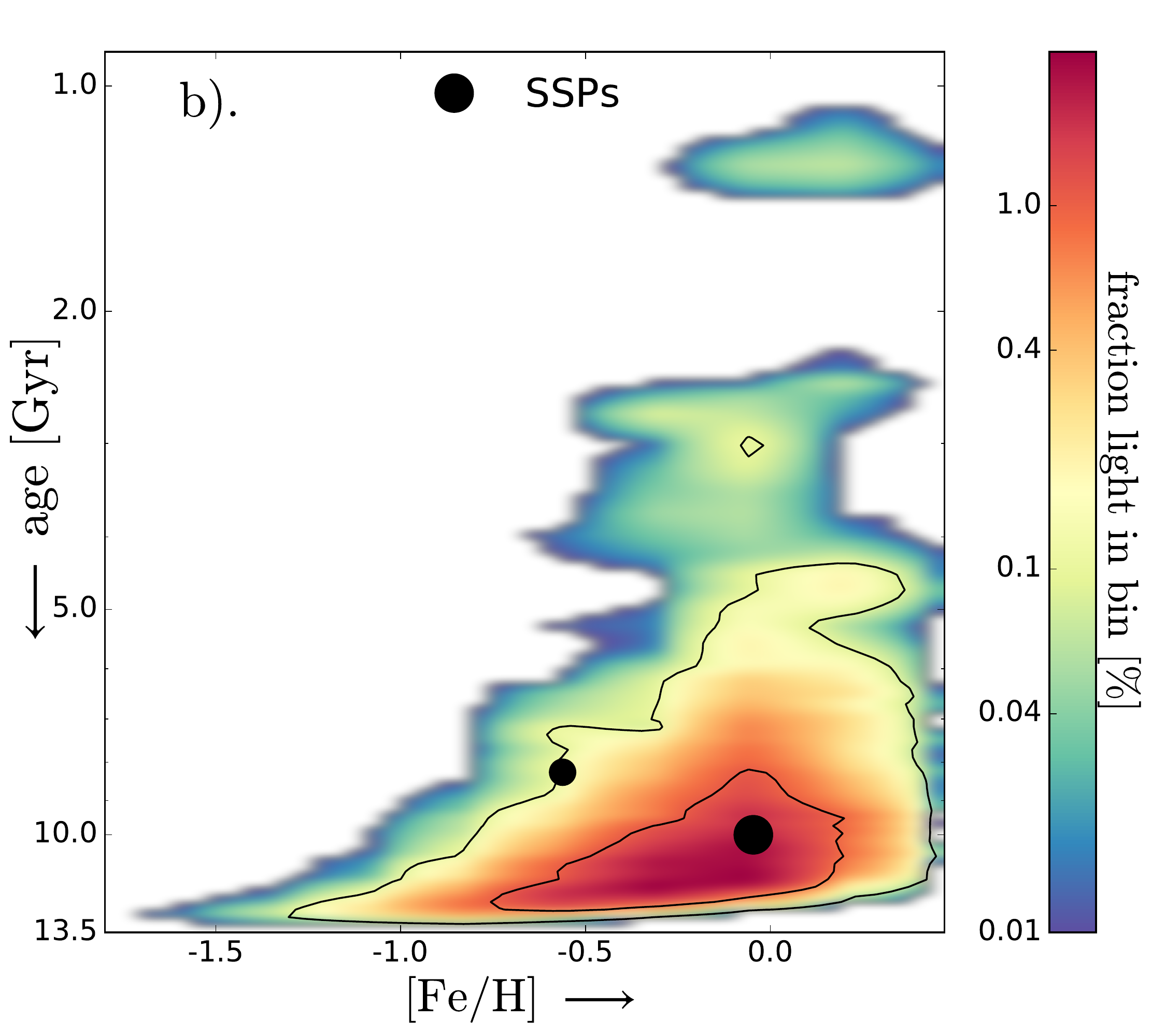}}
	{\includegraphics[width=0.3\textwidth]{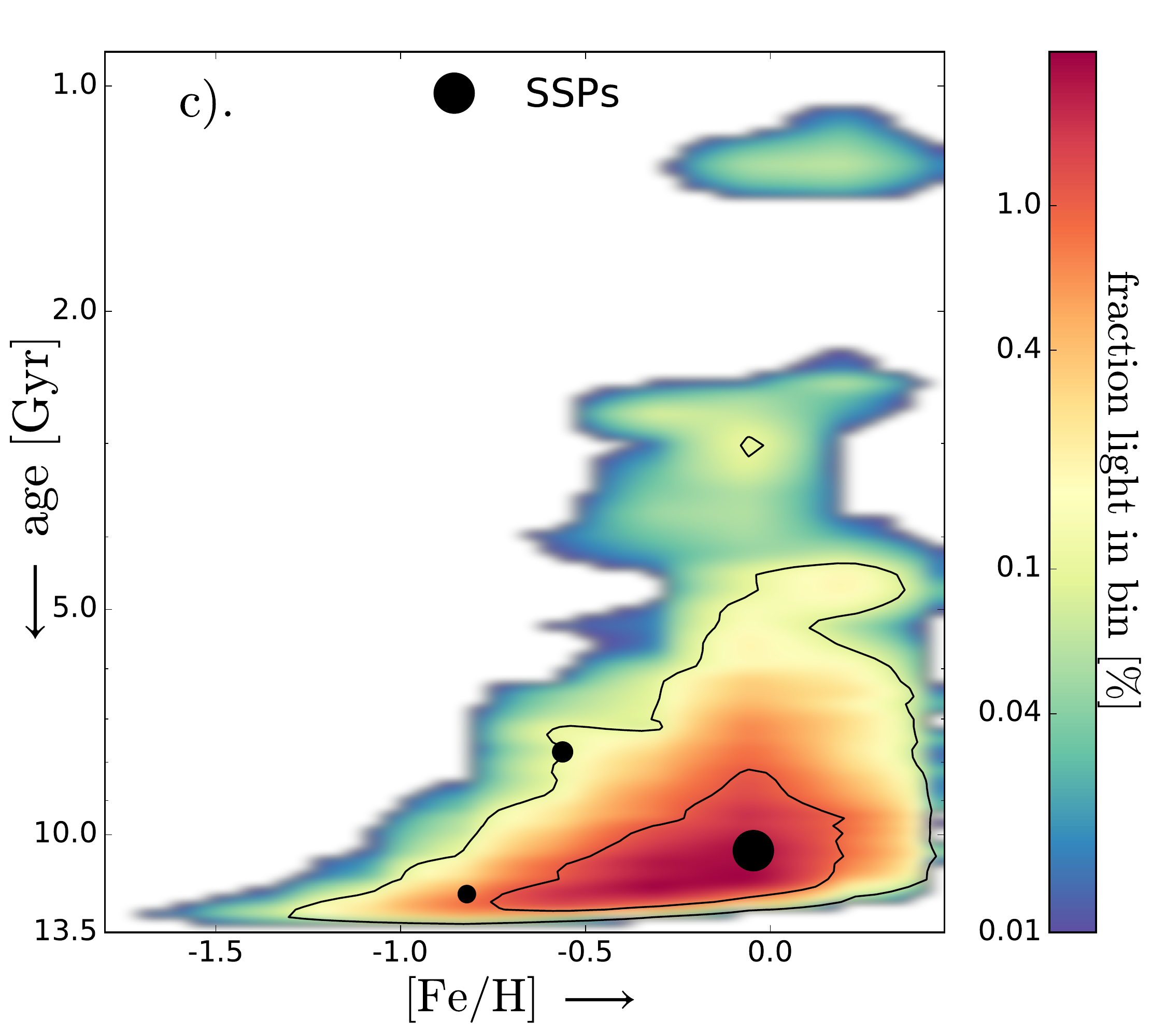}}
	{\includegraphics[width=0.3\textwidth]{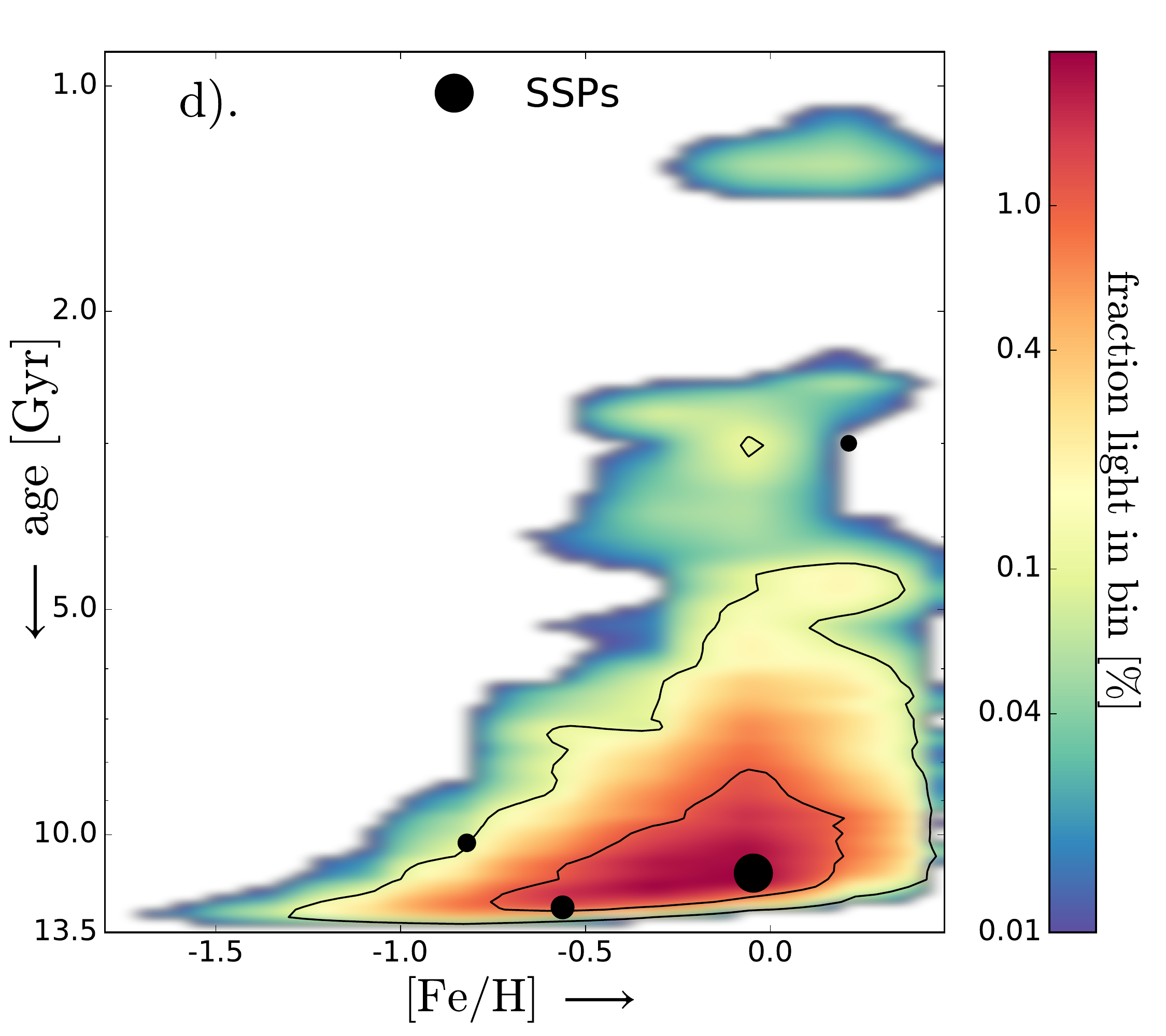}}
	{\includegraphics[width=0.3\textwidth]{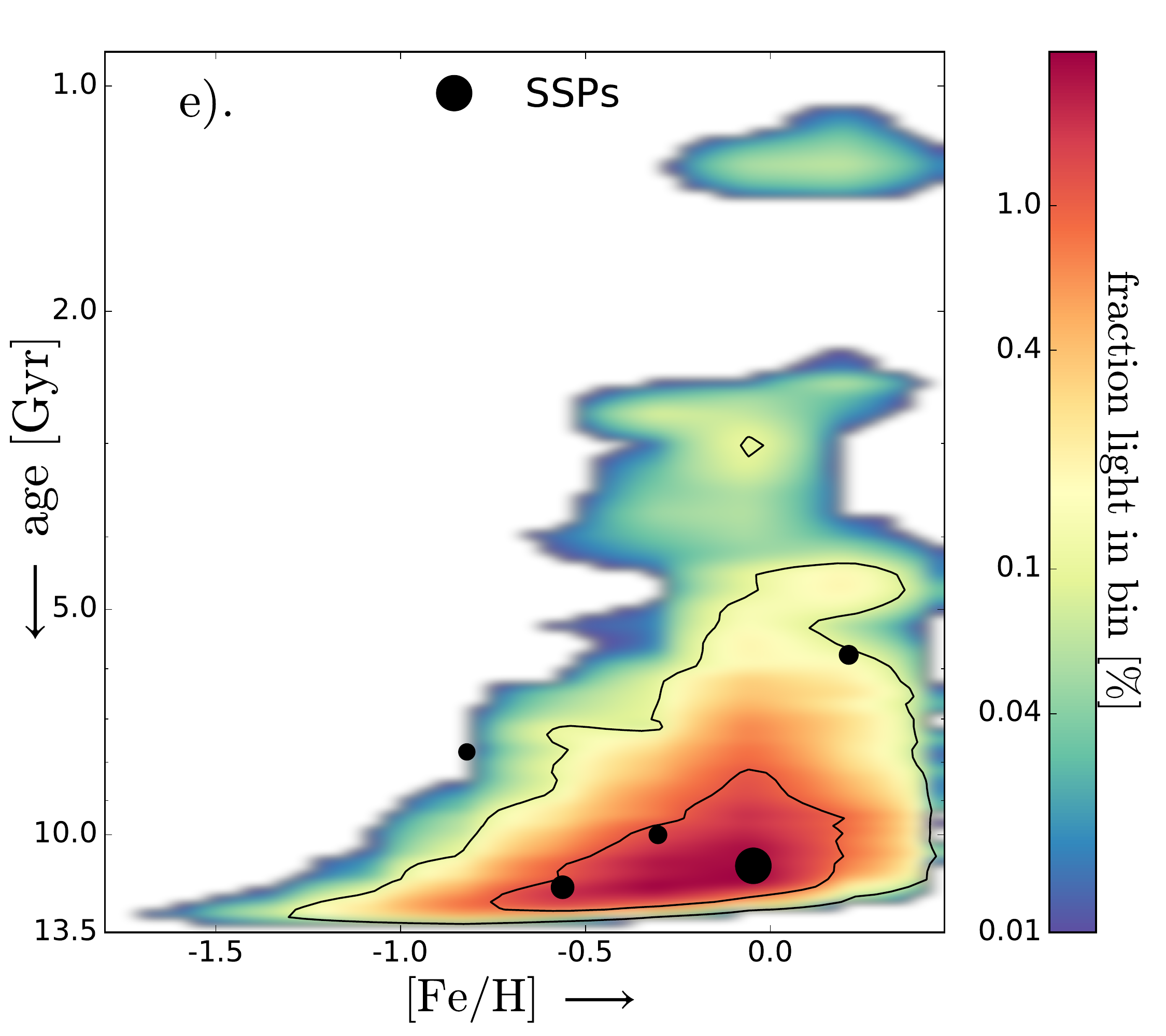}}
	{\includegraphics[width=0.3\textwidth]{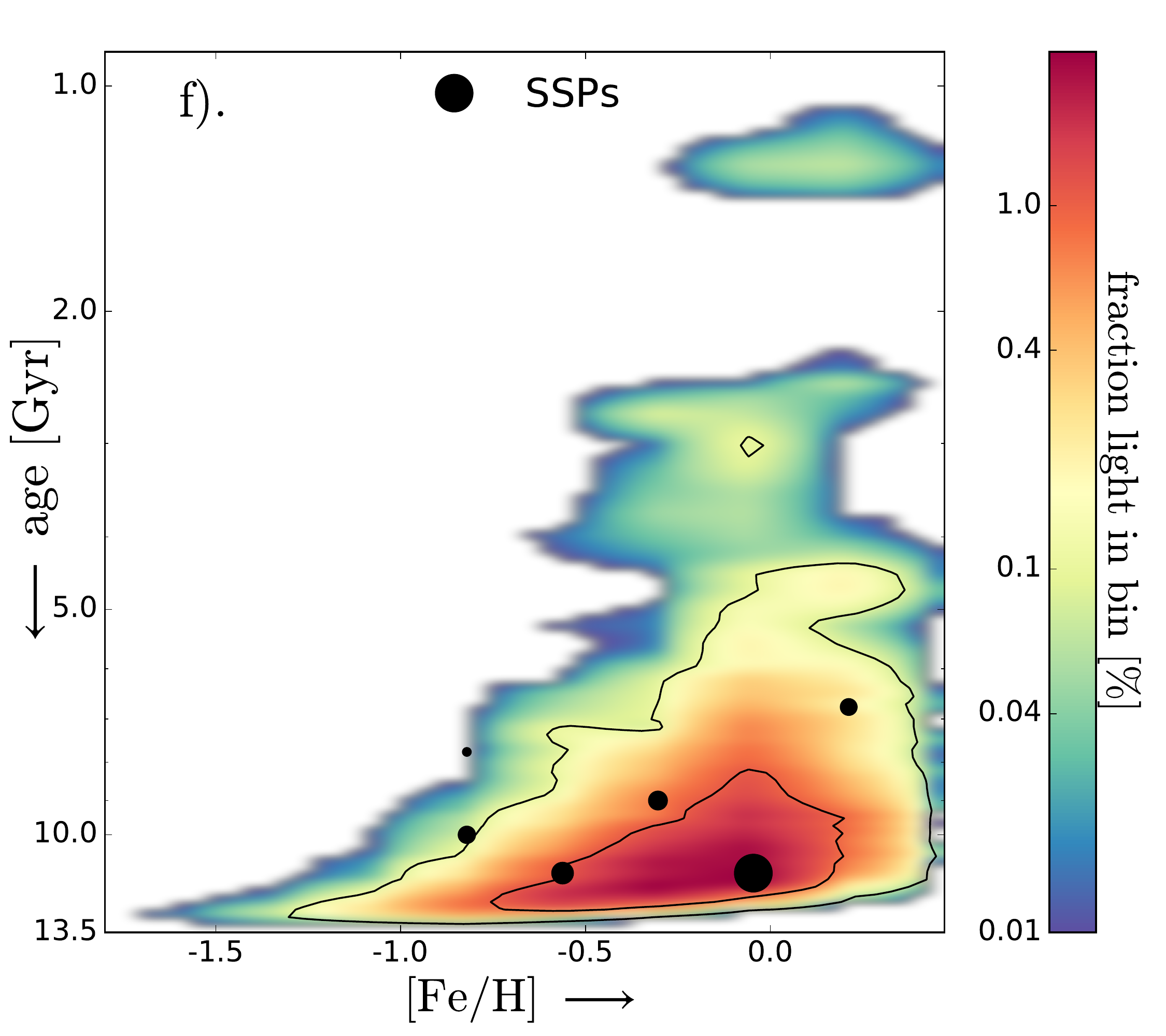}}
	\caption{Reconstructed ages and metallicities of the SSPs used to fit \texttt{CSP150-150}. The colour coding and contours correspond to the SFH of Figure \ref{SFHs}a and the black dots represent the ages and metallicities of the SSPs fitted to the spectrum. The bigger the dot, the higher the relative contribution of that SSP to the integrated spectrum. Different panels correspond to a different number of allowed SSPs in the fit: (a). 1 SSP, (b). 2 SSPs, (c). 3 SSPs, (d). 4 SSPs, (e). 5 SSPs, and (f). 6 SSPs.}
	\label{SFH150}
\end{figure*}

In Figure \ref{SFH150} we show the reconstructed ages and metallicities of the SSPs used to fit \texttt{CSP150-150}\footnote{Here we use \texttt{CSP150-150} to refer to the version of \texttt{CSP150} with SNR = 150.} and where we allow for 1-6 SSPs. For the analysis with one SSP, the age of the SSP is clearly biased towards younger ages. Light-weighted SSP-equivalent ages are known to be biased towards younger ages and to underestimate the age of a CSP \citep{Trager2000, Serra2007, TragerSomerville2009}, and we expect that the bias in Figure \ref{SFH150}a is similar to this more generally known bias. However, for two or more SSPs the most prominent SSP (biggest dot) falls well within the central contour of the SFH of \texttt{CSP150}. Although the ages of the different SSPs in one fit can be quite similar, it is interesting to see that the model almost never selects two SSPs with the same metallicity. This may be related to the coarser sampling of the metallicity grid, and SSPs at different metallicities may also be used in part to absorb the effect of SSPs at different ages due to the age-metallicity degeneracy.

\subsection{Reconstructed IMFs}
As a next step we consider the reconstruction of the IMF for \texttt{CSP150-150}. To determine the reconstructed IMF we first sample the IMF prior parameters at fixed ages and metallicities of the SSPs\footnote{The ages and metallicities of the SSPs are determined with the parameterized version of the model in a previous step.}. Then we select the maximum-a-posteriori (MAP) set of IMF prior parameters to construct a prior on the weights and we use this prior to calculate the most probable weights. Finally, using Equation \ref{eq:IMFweights1}, we convert the most probable weights into a most probable reconstructed IMF. The error bars on the reconstructed IMF are derived on the basis of the posterior sample of the weights.

\begin{figure}
	\centering
		{\includegraphics[width=0.9\columnwidth]{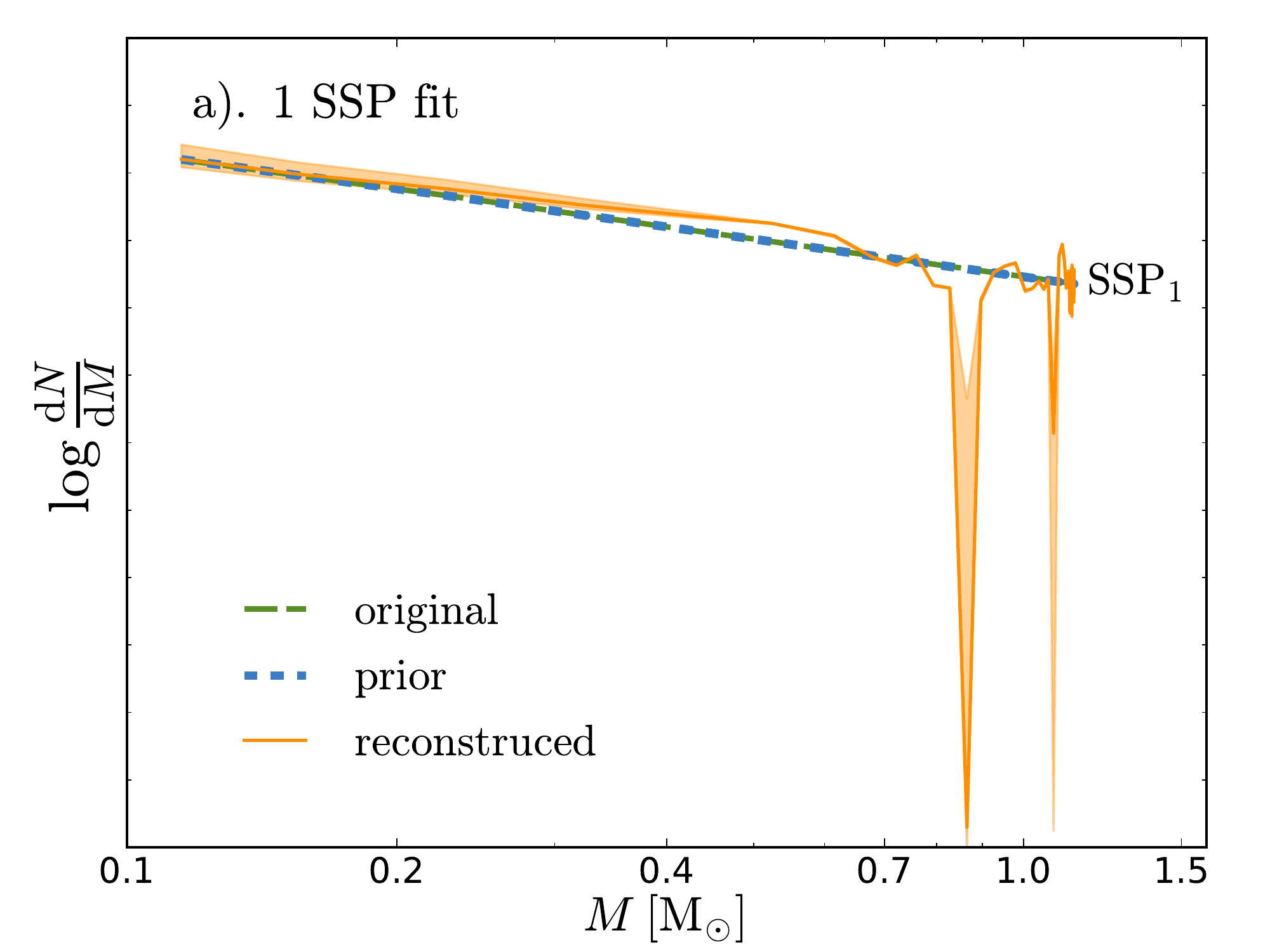}}
		{\includegraphics[width=0.9\columnwidth]{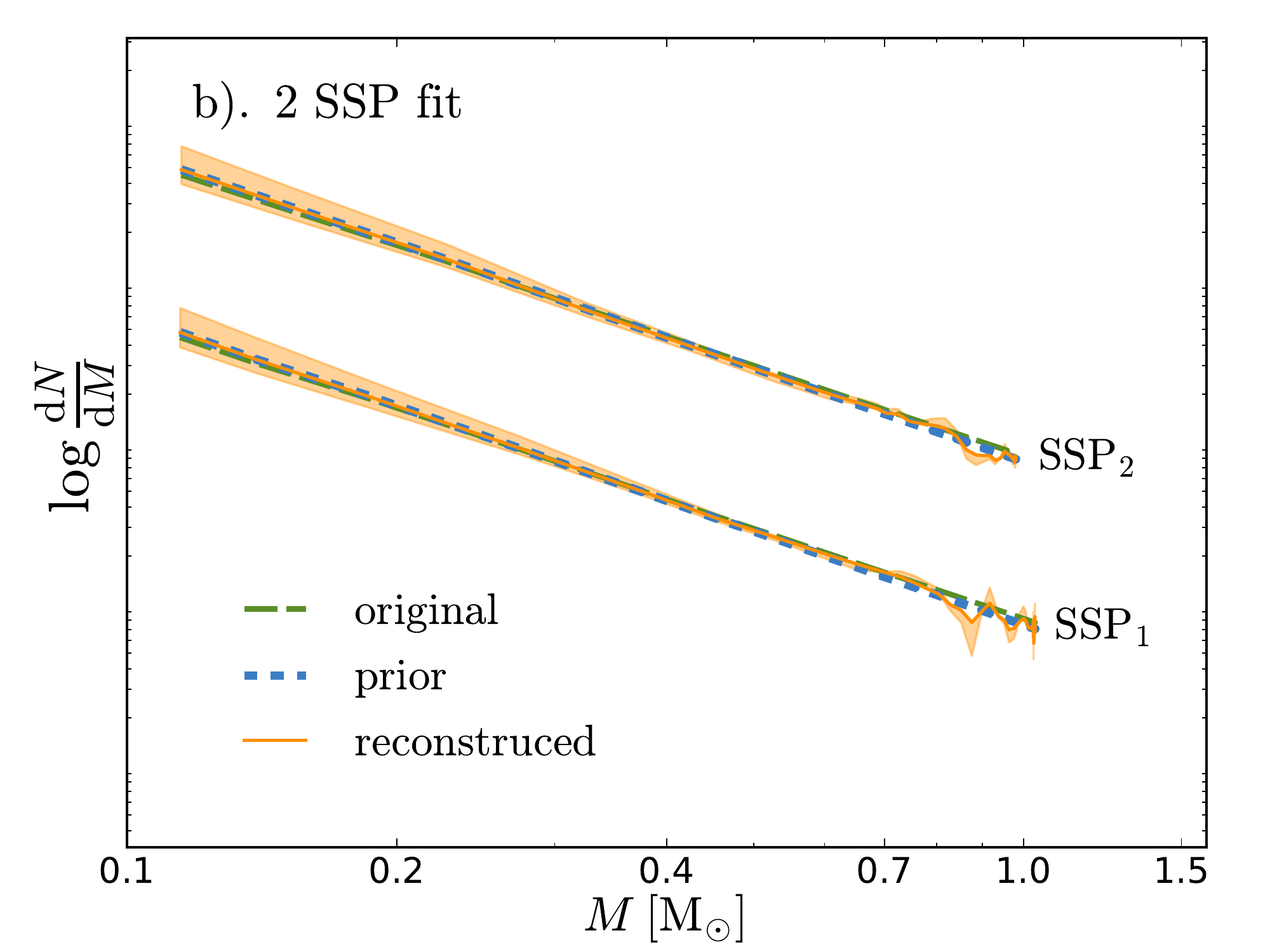}}
		{\includegraphics[width=0.9\columnwidth]{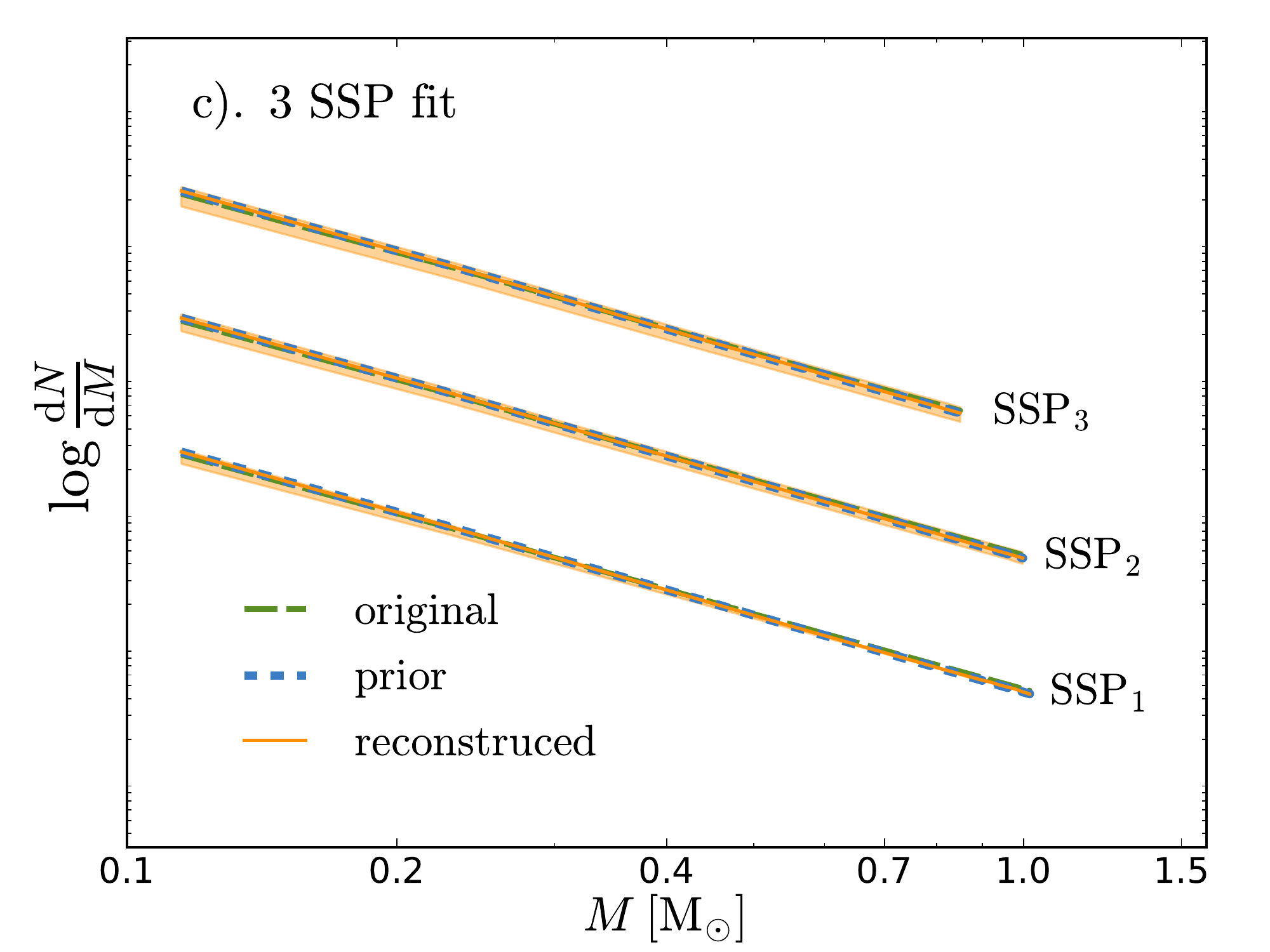}}
	\caption{Reconstructed IMF of \texttt{CSP150-150}. The \textit{green-long-dashed} line represent the original IMF, the \textit{blue-short-dashed} line the prior IMF and the \textit{orange} line the most probable reconstructed IMF. The \textit{shaded-orange} region represents the one sigma confidence interval of the reconstructed IMF and is derived from the distribution of most probable weights of the posterior sample. In panel (a) we allow for one SSP in the fit, in panel (b) for two SSPs and in panel (c) for three SSPs.}
	\label{IMF150}
\end{figure}

The reconstructed IMF of \texttt{CSP150-150} is shown in Figure \ref{IMF150} for models with $N = 1,2,3$ SSPs. For the analysis with one SSP, the most probable IMF deviates significantly from the prior IMF and the original IMF. This already shows that, with one SSP, a single power law IMF is not able to provide a fit to the data that is consistent with the specified noise level. If such a model would have been able to fit the data there would be no need for the model to deviate from the prior. Of course it makes sense that this model does not fit the data since our spectrum in reality represents a CSP with an extended SFH. 

For two SSPs the reconstructed IMF is already much closer to the prior and the original IMF, although there are still some wiggles at the high-mass end. When we use three (or more) SSPs in the fit, the prior IMF, the original IMF and the reconstructed IMF become very similar. On the basis of this similarity we might conclude that three SSPs are sufficient to fit the spectrum of \texttt{CSP150-150}. According to Figure \ref{SSPevidence}, for this particular case the evidence continues to increase a little bit more for $n_{\mathrm{SSPs}} > 3$ but the difference in log evidence between $N = 3$ and $N = 4$ SSPs is 4.9 which according to \cite{Jeffreys1961} is not considered to be substantial evidence in favour of $N = 4$ SSPs. Three SSPs therefore seem to be sufficient here.

\subsection{Reconstructed spectra}
Finally, we consider the reconstructed spectra of \texttt{CSP150-150}. To determine the reconstructed spectrum we start with the MAP values of the sampled non-linear parameters and we use these values to calculate the most probable weights. Then we multiply the stellar-template matrix with the most probable weights to obtain the reconstructed MAP spectrum.

\begin{figure*}
	\centering
	{\includegraphics[width=0.45\textwidth]{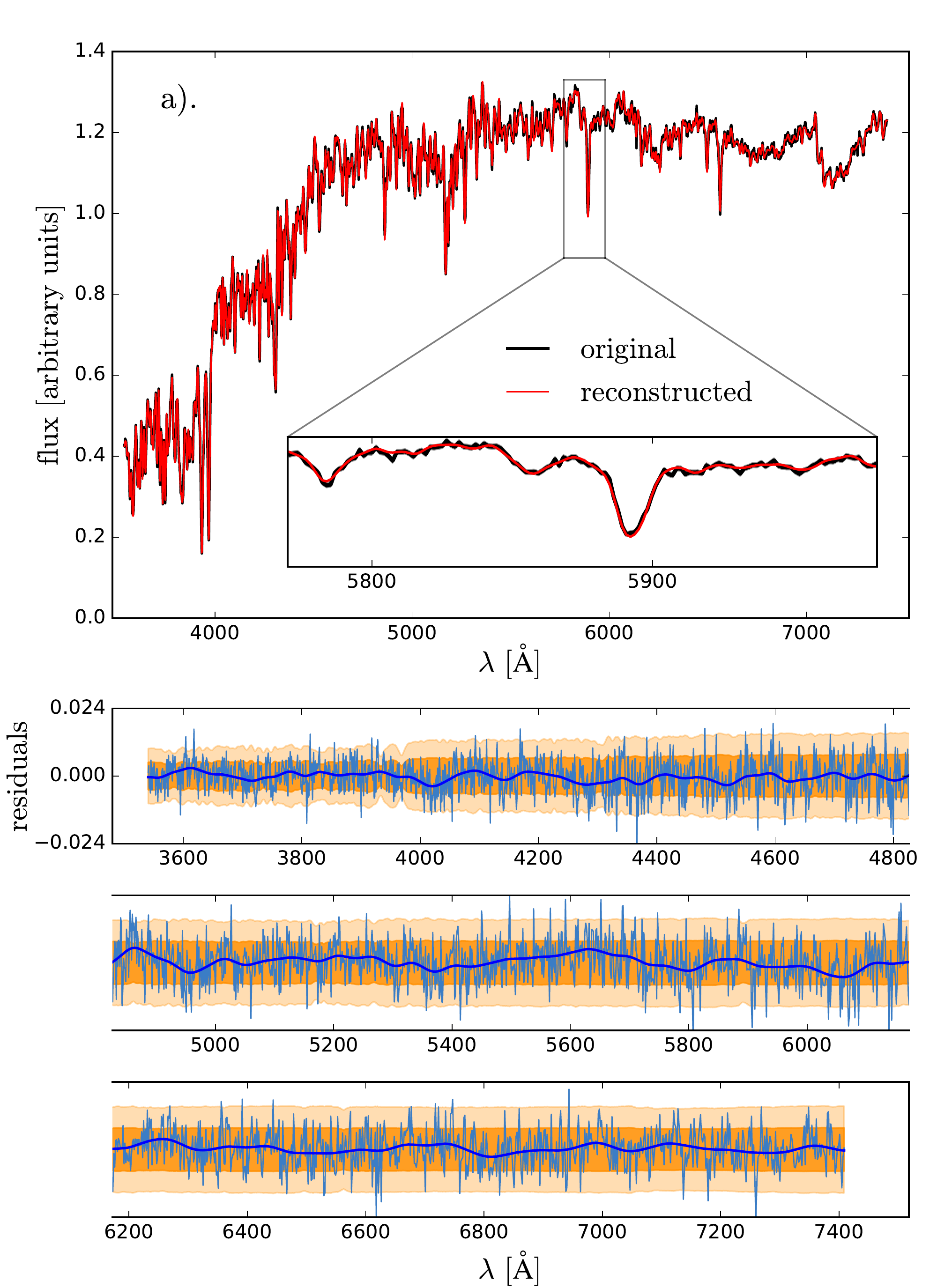}}
		{\includegraphics[width=0.45\textwidth]{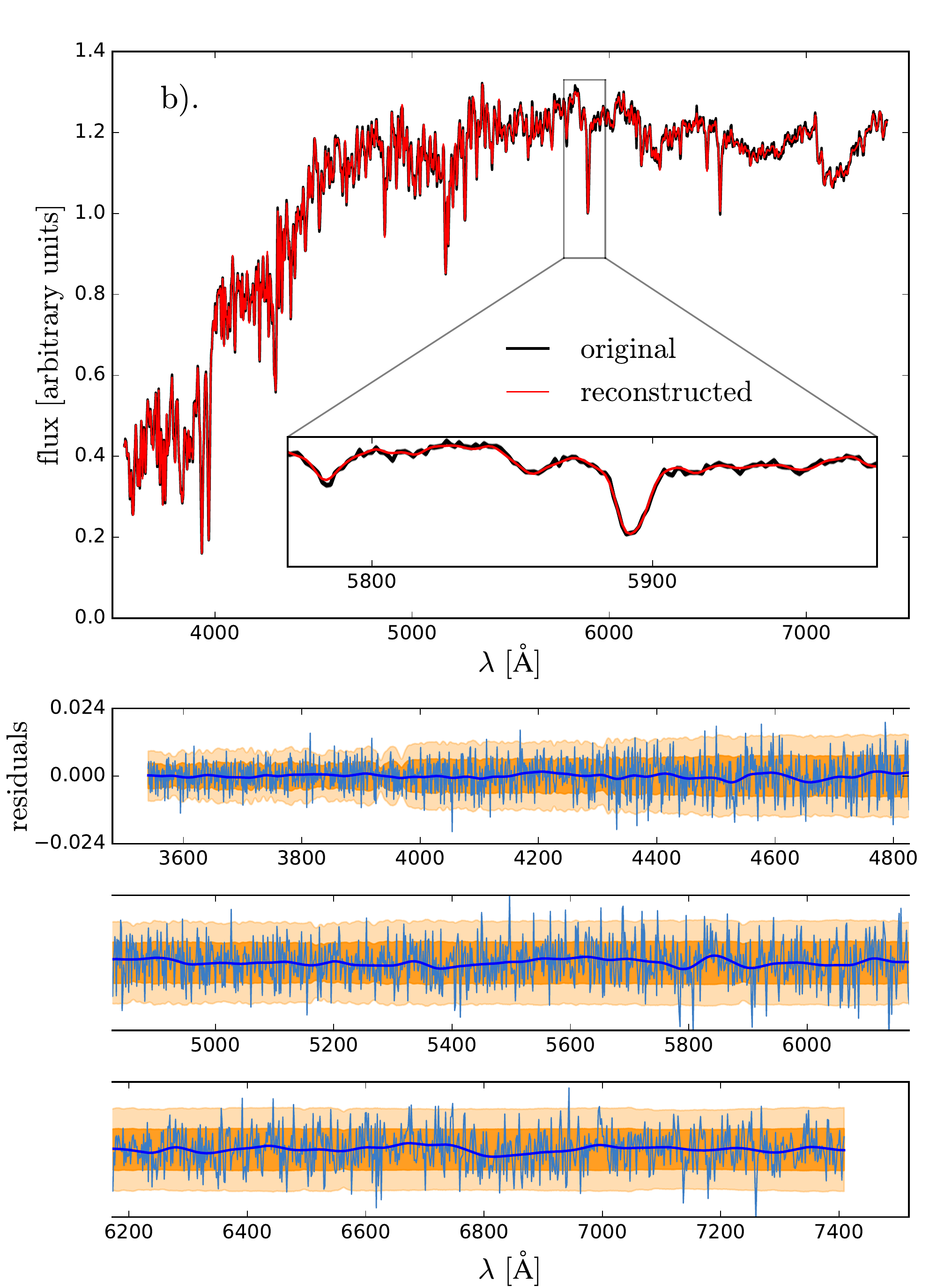}}
	\caption{Reconstructed spectrum of \texttt{CSP150-150}. The \textit{black} line represents the original spectrum and the \textit{red} line the reconstructed spectrum. In panel (a) we use one SSP for the fit and in panel (b) we use five SSPs. The bottom plots represent the residuals in \textit{blue}. Also shown is a smoothed version of the residuals with the \textit{thick-blue} lines. The \textit{shaded-orange} region represent the one and two sigma errors of the spectrum and are derived from the covariance matrix $\mathbfss{C}_{\mathrm{D}}$.}
	\label{spectra150}
\end{figure*}

The reconstructed MAP spectrum of \texttt{CSP150-150} is shown in Figure \ref{spectra150}. In the first panel, we use only one SSP for the fit whereas in the second panel we use five SSPs which is the maximum that we find in the evidence as a function of the number of SSPs. Looking at the reconstructed spectra for one and for five SSPs, at first glance there does not seem to be a difference. However, if one carefully looks at the residuals there appears to be a wave-pattern in the residuals when we use one SSP. If we use five SSPs, this pattern is far less prominent. To visualize this noise-pattern we also show a smoothed version of the residuals in Figure \ref{spectra150}.

\subsection{Reconstructing the variable IMF}
As discussed in Section \ref{sec:spectraConstruction}, the IMF of our CSP mock spectra varies as a function of velocity dispersion. All of our mock spectra have a single power law IMF for which the IMF slope is given by Equation \ref{eq:SpinielloRelation}. In this section, we try to reconstruct the underlying trend of the IMF from our mock spectra for different SNRs.

We analyse all of our mock spectra using the sampling strategy discussed in Section \ref{sec:samplingStrategy}. First we derive the most probable ages and metallicities as well as the velocity dispersion with the parameterized version of the model. We then fix these parameters and run the full version of the model to calculate the evidence and obtain posterior samples of the IMF prior parameters, the weights and the covariance parameter $b$. For each of the mock spectra, we repeat the analysis with $N = 1-6$ SSPs, and we select the number of SSPs that maximizes the evidence.

\begin{table*}
	\centering
	\caption{Reconstructed parameters of CSP mock spectra for different SNRs. The different columns of this table give the name of the CSP, the number of SSPs that maximizes the evidence, the ages of these SSPs, the metallicities of these SSPs, the input value of the IMF slope that was used to create the spectrum, the reconstructed IMF slope, the reconstructed velocity dispersion, and the reconstructed covariance parameter. Errors on the IMF slope correspond to the 16th and 84th percentile of the posterior distribution, errors for $\sigma$ and $b_{\mathrm{cov}}$ correspond to the standard deviation of the posterior sample. Note that with \texttt{CSP150-75} we refer to \texttt{CSP150} with SNR = 75.}
	\label{tab:results}
	\begin{tabular}{cccccccc} 
		\hline
		name & $n_{\mathrm{SSPs}}$ & SSP & SSP & $\alpha_{\mathrm{in}}$ & $\alpha_{\mathrm{rec}}$ & $\sigma_{\mathrm{rec}}$ & $b_{\mathrm{cov,rec}}$\\
 & &	 ages [Gyr] & [Fe/H]s\\
		\hline
		\rule{0pt}{3ex}
		\texttt{CSP150-75} & 3 & 10.5, 9.5, 9.0 & -0.05, -0.56, -1.08 & 1.84 & $1.76^{+0.12}_{-0.13}$ & $150.0\pm 1.1$ & $0.02\pm 0.01$\\
		\rule{0pt}{3ex}
		\texttt{CSP190-75} & 3 & 9.8, 12.0, 11.0 & -0.05, -0.56, -0.21 & 2.08 & $2.08^{+0.11}_{-0.11}$ & $190.9\pm 1.2$ & $0.02\pm 0.01$\\
		\rule{0pt}{3ex}
		\texttt{CSP230-75} & 4 & 11.5, 11.8, 8.3, 10.8 & -0.05, -0.21, -0.30, -1.08 & 2.27 & $2.28^{+0.10}_{-0.10}$ & $233.3\pm 1.4$ & $0.02\pm 0.01$\\
		\rule{0pt}{3ex}
		\texttt{CSP270-75} & 3 & 12.3, 11.0, 7.5 & -0.05, 0.21, -0.56 & 2.43 & $2.39^{+0.13}_{-0.11}$ & $269.8\pm 1.5$ & $0.01\pm 0.01$\\
		\rule{0pt}{3ex}
		\texttt{CSP310-75} & 2 & 9.8, 12.0 & 0.21, -0.30 & 2.57 & $2.50^{+0.10}_{-0.12}$ & $312.6\pm 2.1$ & $0.01\pm 0.01$\\
		\hline
		\rule{0pt}{3ex}
		\texttt{CSP150-150} & 5 & 11.0, 10.0, 5.8, 11.8, 7.8 & -0.05, -0.30, 0.21, -0.56, -0.82 & 1.84 & $1.88^{+0.07}_{-0.07}$ & $150.0\pm 0.5$ & $0.01\pm 0.01$\\
		\rule{0pt}{3ex}
		\texttt{CSP190-150} & 4 & 13.0, 8.0, 10.0, 13.3 & -0.05, 0.21, -0.56, -0.82 & 2.08 & $2.06^{+0.04}_{-0.05}$ & $188.8\pm 0.6$ & $0.01\pm 0.01$\\
		\rule{0pt}{3ex}
		\texttt{CSP230-150} & 5 & 11.0, 11.5, 12.0, 4.5, 9.8 & -0.05, 0.21, -0.56, 0.21, -1.08 & 2.27 & $2.26^{+0.06}_{-0.06}$ & $229.8\pm 0.7$ & $0.02\pm 0.01$\\
		\rule{0pt}{3ex}
		\texttt{CSP270-150} & 4 & 10.5, 10.8, 11.0, 12.0 & 0.21, -0.05, -0.56, 0.47 & 2.43 & $2.43^{+0.04}_{-0.05}$ & $268.7\pm 0.8$ & $0.01\pm 0.01$\\
		\rule{0pt}{3ex}
		\texttt{CSP310-150} & 3 & 10.0, 12.3, 11.5 & 0.21, -0.05, -0.56 & 2.57 & $2.59^{+0.04}_{-0.04}$ & $310.1\pm 1.0$ & $0.02\pm 0.01$\\
		\hline
		\rule{0pt}{3ex}
		\texttt{CSP150-300} & 6 & 11.5, 8.3, 9.8, 10.8 & -0.30, -0.05, 0.21, -0.56 & 1.84 & $1.80^{+0.03}_{-0.03}$ & $150.5\pm 0.3$ & $0.02\pm 0.02$\\
		\rule{0pt}{3ex}
		 &  & 9.5, 4.0 & -1.08, -1.08\\
		\rule{0pt}{3ex}
		\texttt{CSP190-300} & 5 & 10.5, 10.3, 10.0, 12.0, 11.5 & -0.05, 0.21, -0.30, -0.56, -1.6 & 2.08 & $2.12^{+0.02}_{-0.02}$ & $190.2\pm 0.3$ & $0.02\pm 0.01$\\
		\rule{0pt}{3ex}
		\texttt{CSP230-300} & 4 & 10.0, 11.0, 11.8, 7.5 & 0.21, -0.05, -0.30, -0.56 & 2.27 & $2.32^{+0.02}_{-0.03}$ & $229.7\pm 0.4$ & $0.02\pm 0.01$\\
		\rule{0pt}{3ex}
		\texttt{CSP270-300} & 5 & 10.5, 12.3, 11.8, 6.8, 1.8 & -0.05, 0.21, -0.56, 0.47, -0.05 & 2.43 & $2.46^{+0.03}_{-0.02}$ & $270.5\pm 0.4$ & $0.01\pm 0.01$\\
		\rule{0pt}{3ex}
		\texttt{CSP310-300} & 5 & 11.3, 11.5, 8.0, 13.0, 12.3 & -0.05, 0.21, 0.46, -0.30, -0.82 & 2.57 & $2.60^{+0.03}_{-0.02}$ & $310.3\pm 0.5$ & $0.02\pm 0.02$\\
		\hline
	\end{tabular}
\end{table*}

The results of our analysis are given in Table \ref{tab:results}. This table summarizes the number of SSPs that maximizes the evidence, the ages and metallicities of these SSPs, the reconstructed IMF slope, the reconstructed velocity dispersion and the reconstructed covariance parameter for each of our CSPs. The results in Table \ref{tab:results} confirm our results in Section \ref{sec:reqNssps}: spectra with higher SNRs and spectra with more extended SFHs in general require more SSPs. Since the SFHs of the CSPs are extended, we cannot directly compare the reconstructed ages and metallicities of the fitted SSPs. However, if we consider the ages and metallicities of the SSPs in Figure \ref{SFHs}, they seem to cover the distribution function nicely. We want to emphasize that the aim of including multiple SSPs in the fit is not to reconstruct the SFH perfectly but to ensure that the determination of the IMF is not biased as a consequence of assuming only one SSP. In that respect, the age-metallicity distribution can be regarded for the purpose of this paper as nuisance parameters that are marginalized over. However, our analysis shows that even high-SNR spectra typically contain information of at most half a dozen distinct stellar-population regions in their space of ages and metallicities. This makes full-Bayesian modelling of these spectra to infer their IMFs and SFHs a tractable problem on present-day computers.

\begin{figure}
	\centering
		{\includegraphics[width=0.9\columnwidth]{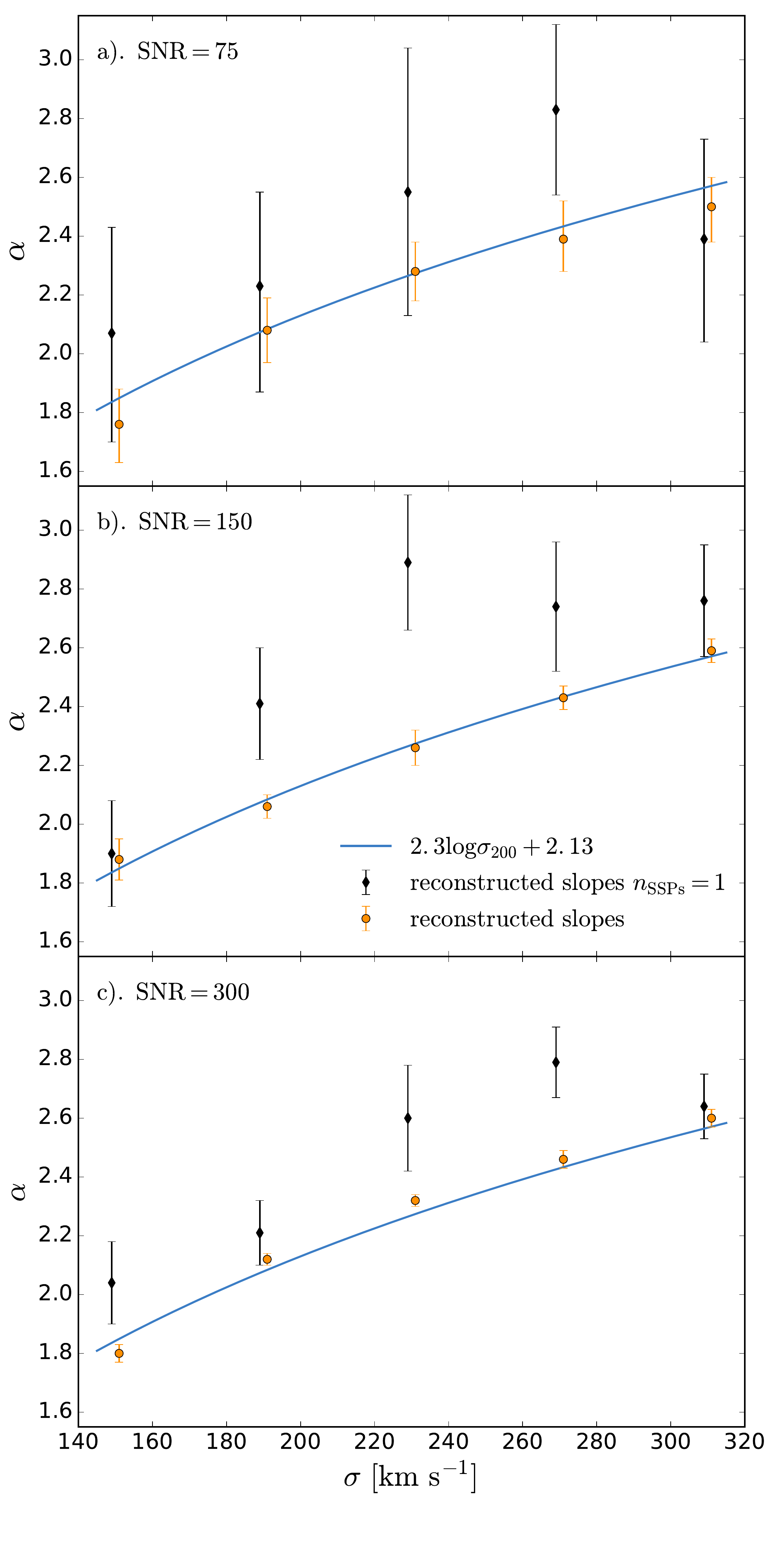}}
	\caption{Reconstructed IMF slopes versus (true) velocity dispersion for CSP mock spectra with different SNRs. Data points represented by \textit{black diamonds} are derived by using one SSP whereas for the \textit{orange circles} we used the number of SSPs that maximizes the evidence. The error bars correspond to the 16th and 84th percentile of the posterior distribution. The blue line corresponds to the IMF slope--velocity dispersion relation of Equation \ref{eq:SpinielloRelation}.}
	\label{SpRecAll}
\end{figure}

The reconstructed velocity dispersions in Table \ref{tab:results} are consistent with their input values. We have parameterized the extra covariance term by $b_{\mathrm{cov}}$, which is the extra (constant) covariance in terms of the median of the original covariance matrix, i.e.
\begin{equation}
b = b_{\mathrm{cov}}\cdot \mathrm{median}\left( \mathbfss{C}_{\mathrm{D}} \right).
\end{equation}
From the results in Table \ref{tab:results} we can see that $b_{\mathrm{cov}}$ is at most 0.02 so that the additional covariance is in general not more than 2\% of the median of the original covariance matrix. The additional covariance term does therefore not seem to play an important role here, but it might become more important if we deal with real data where we have more systematic uncertainties.

The reconstructed IMF slopes are consistent with the input values. In Figure \ref{SpRecAll} we show the reconstructed IMF slopes (orange data points) together with the IMF slope--velocity dispersion relation specified in Equation \ref{eq:SpinielloRelation} for SNRs of 75, 150 and 300, respectively. These IMF slopes are derived by using the number of SSPs that maximizes the evidence. The results show that for all SNRs we are able to correctly reconstruct the IMF slope--velocity dispersion relation. As expected, for lower SNRs the scatter in this relation increases. For SNR=300, the error bars seem to be slightly underestimated, or the results are slightly biased, but the overall trend is correct and the blue line in Figure \ref{SpRecAll}c would still fall within the two sigma error bars of the data points.

\subsubsection{IMF slope as a function of $n_{\mathrm{SSPs}}$}
We have seen that the evidence prefers fits with multiple SSPs over fits with only one SSP for all simulated spectra with realistic SFHs. Here we investigate in more detail if the determination of the IMF slope is affected by the number of SSPs that we use in our fit.

\begin{figure}
	\centering
		{\includegraphics[width=0.9\columnwidth]{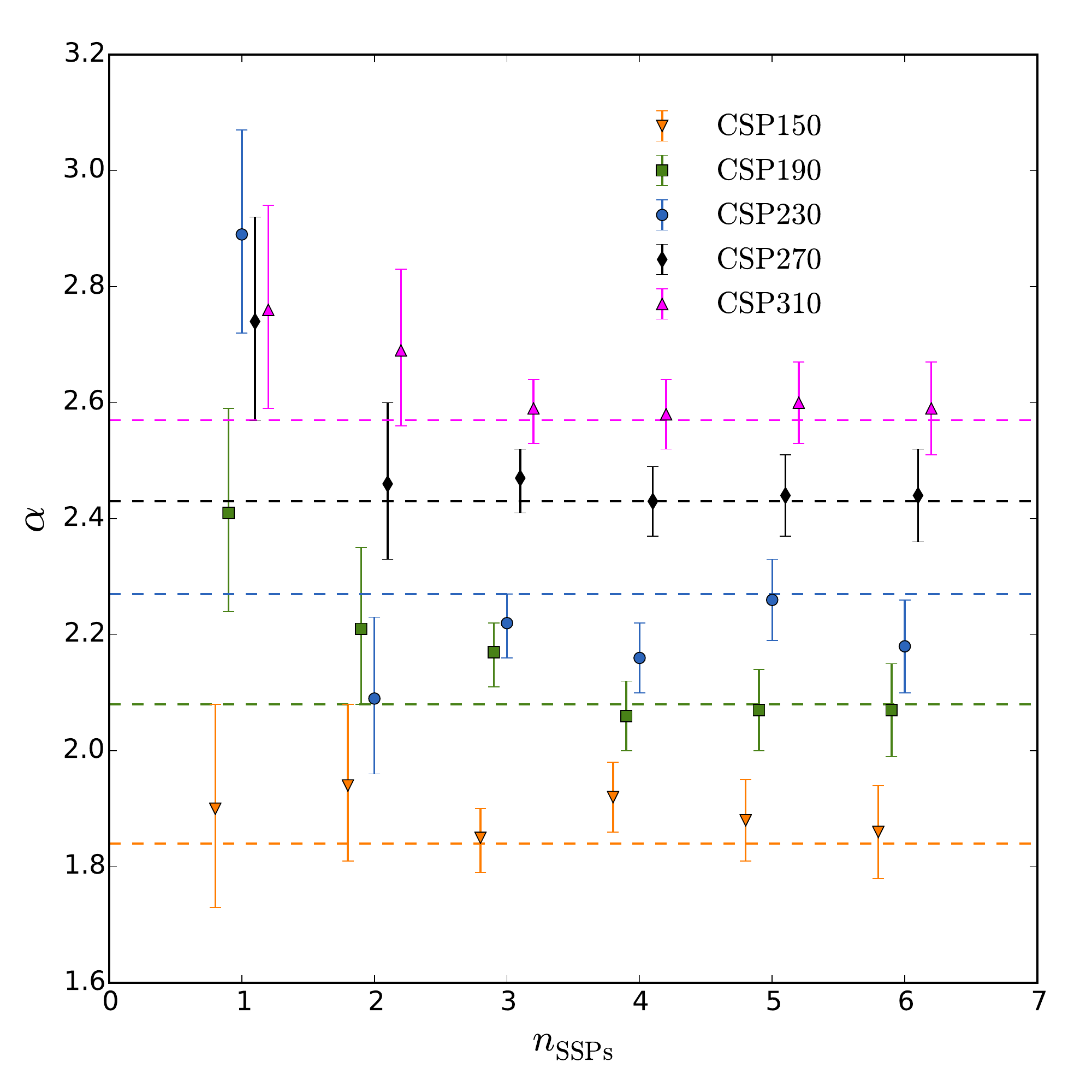}}
	\caption{Reconstructed IMF slopes as a function of the number of SSPs used in the fit for all mock spectra with SNR=150. The dashed lines correspond to the input values of the IMF slope used to create the mock spectrum. Different velocity dispersions are slightly offset to allow the reader to distinguish them.}
	\label{IMFvsnSSPs150}
\end{figure}

The black data points in Figure \ref{SpRecAll} show the IMF-slope--velocity dispersion relation that we find if we use only one SSP. Except for \texttt{CSP310-75}, the IMF slopes inferred by using only one SSP are biased to a higher value. The overall trend of an IMF slope that increases as a function of velocity dispersion is still recovered, however.

In Figure \ref{IMFvsnSSPs150} we show the reconstructed IMF slope as a function of the number of SSPs used in the fit for all mock spectra with SNR=150. As we have seen in Figure \ref{SpRecAll}, when we use only one SSP the reconstructed IMF slopes are biased and the slopes that we derive are in general too high. For two SSPs there is still some bias, but for nearly all CSPs the input value of the IMF slope lies within the one sigma confidence interval of the reconstructed IMF slope. If we use three or more SSPs, the reconstructed IMF slopes converge to their true values, and the error bars on the reconstructed IMF slopes become approximately constant for more than three SSPs. So over a realistic range of SFHs, velocity dispersion (150-310 km s$^{-1}$) and SNRs (75-300), in general $N=2-3$ SSPs are sufficient to recover the IMF without significant bias.

Given the wavelength region and the resolution of the data that we consider, we conclude that it should be sufficient to model CSPs with SFHs similar to the ones in Figure \ref{SFHs} with $\sim$3 SSPs. If we use only one SSP, this might significantly bias our reconstructed IMF slope. Although we still recover the trend, it is shifted upwards, indicating an overall overestimation of the IMF slope.

\section{Summary and discussion}
We have extended the hierarchical Bayesian framework of DTK16 to include multiple SSPs, a multiplicative polynomial and an extra covariance term. In addition, we have developed a parameterized version of the code which is much faster than the full Bayesian version. We use that version of the code to determine the velocity dispersion and the ages and metallicities of the SSPs that we fit to a spectrum. To reduce the number of templates in our model, we have implemented a binning procedure where we combine different isochrone stars.

We have applied the updated version of our model to a set of CSP mock spectra with different SNRs, stellar velocity dispersions and age-metallicity distributions. Based on the evidence, we determine the number of SSPs required to fit the mock spectra. The higher the SNR and the more extended the SFH of a CSP, the more SSPs are required to fit the spectrum correctly. As a rule of thumb, for SNR=75 there is strong evidence to include 2 SSPs in the fit, for SNR=150 there is strong evidence to include 2-3 SSPs in the fit and for SNR=300 there is strong evidence to include 5 SSPs in the fit. However, for removing the bias in the IMF slope $N = 2-3$ seems sufficient even for the high SNR = 300 cases.

Our CSP mock spectra have a single power law IMF with a slope that varies as a function of the present-day velocity dispersion. The assumed IMF slope--velocity dispersion relation is purely driven by observational results. However, we do not expect this to be a physically realistic model since most of the stars in these galaxies formed long ago when the galaxy properties were quite different. Therefore, physically we expect that a variable IMF is driven by different local properties of the interstellar medium (ISM) at the time that the stars are born. These local properties of the ISM during star formation may typically be different for galaxies with different velocity dispersions/masses, which would then explain the observational results. In this work we use these observational results as a simple parameterization of the IMF but in principle within cosmological simulations one can relate the IMF to local conditions at the sites of star formation \citep[see][]{Blancato2016}.

We show that we can reconstruct the underlying IMF slope--velocity dispersion relation for all the SNRs we consider. For this analysis, we fit the mock spectra with multiple SSPs and choose the number of SSPs that maximizes the Bayesian evidence. If we use only one SSP in our fit, our results show that the IMF slope that we determine is likely to be biased upward, although the trend remains. For three SSPs or more the reconstructed IMF slope appears to converge to a stable value but most of the bias is already removed by using two SSPs instead of one. In fact, \cite{Conroy2012b} use two SSP components in their fits, although the age of the second component is fixed to 3 Gyr. \cite{MartinNavarro2015} test the robustness of their inferred IMF gradient for NGC4552 by also modelling it with two SSPs instead of one and conclude that it makes no difference to the inferred gradient. Indeed, the inferred gradient remains if two SSPs are used instead of one, but the inferred IMF slopes appear to be lower, in particular for the outer radii. Our results suggest that IMF slopes derived on the basis of single SSP fits might be biased to a value that is too high if the SFH of the studied galaxy is in reality extended. Since high IMF slopes imply a higher stellar mass, this bias might help to explain discrepancies between stellar masses and masses derived through lensing and/or dynamics. However, despite this bias, any trends that are discovered appear to be real.

Although we have implemented an additional (constant) covariance term to absorb systematic uncertainties, this parameter does not seem to be significant here. The fact that our mock spectra have extended SFHs might potentially introduce the need to increase the covariance. However, this effect appears to be mostly absorbed by modelling the spectra with multiple SSPs. Note that if we model our spectra with one SSP only, the additional covariance term ($b_{\mathrm{cov}}$) is in general significantly higher. Since the extended SFHs of our mock spectra are the only source of systematic uncertainty in this work, the fact that $b_{\mathrm{cov}}$ is in general relatively small is to be expected. If we deal with real data which is affected by other potential sources of uncertainty we expect this term to become more important, and potentially have a complex covariance matrix structure (e.g. due to correlations).

We model a CSP with a complex SFH by approximating it as the sum of a given number of SSPs. The aim of this approach is to investigate whether the inferred IMF is affected by the number of SSPs used in the fit. Although our method gives us some constraint on the SFH, it does not allow us to fully reconstruct the SFH of the CSP. One could imagine a data model similar to that present here that allows one to infer the SFH and the IMF simultaneously. In that case, the number of stellar templates in the matrix $\mathbfss{S}$ would increase considerably, and we would need to include a parameterized prior on both the SFH and the IMF to regulate the inversion of Equation \ref{eq:linAssumption}. However, that approach is computationally too expensive at present.

As shown in this work, the CSPs that we consider can effectively be modelled by a combination of a few SSPs, since there is a peak in the evidence as a function of the number of SSPs. Cosmological simulations predict that massive galaxies are made up of a significant fraction of stars that are formed in galaxies other then the main progenitor but are later on accreted through mergers \citep{Naab2009,Porter2014,RodriguezGomez2016}. This is one reason why these galaxies are expected to have composite stellar populations. In that context, it would be interesting to see if the SSPs that we recover as the main building blocks of a galaxy correspond to the most important constituent progenitors in the SAM. If so, this would allow us to put constraints on the merger history of the galaxy.

In this study our focus has been on the IMF. If instead one is interested in the SFH, one could assume an IMF and fill matrix $\mathbfss{S}$ with a set of SSP templates instead of stellar templates. By using a parameterized prior on the SFH, one could then create a framework similar to our model that allows for hierarchical Bayesian inference of the SFH \citep[see also][who use regularized inversion to determine non-parametric SFHs of CSPs]{Ocvirk2006}. 

Another possibility to combine inference of both the SFH and the IMF, would be to partition the three dimensional space of $T_{\mathrm{eff}}$, $\log g$ and $\mathrm{[Fe/H]}$ around the isochrones into small cubes. For each of these cubes, we could create a spectrum by using an interpolator and then fill matrix $\mathbfss{S}$ with these spectra. The prior on the weights that we use to to regulate the inversion of Equation \ref{eq:linAssumption} now depends on both the parameterized SFH prior and the parameterized IMF prior. Although in principle this allows for a combined inference of the SFH and the IMF, this approach will still be very computationally expensive and may also be complicated by possible degeneracies between the parameterized SFH and the parameterized IMF. Note that for example \cite{Spinrad1971} and \cite{Faber1972} already developed population synthesis models that tried to determine individual contributions of stars in the HR diagram. However, these methods combined stars in a rather ad hoc way and had few astrophysical constraints. 

In this work we have shown that, given the wavelength range and resolution of the MILES library, we can reconstruct a variable IMF in CSPs. However, it is important to realize that we have only considered single power law parameterizations of the IMF prior. Based on dynamical constraints, \citet{Lyubenova2016} rule out a single power law IMF for the majority of their sample of 27 ETGs whereas it is found to be consistent with a double power law IMF for most of their sample. We might also consider more realistic parameterizations of the IMF, such as a double power law IMF with a break. If we would then create a set of CSP mock spectra where only the low-mass slope varies as a function of velocity dispersion, it could become much harder to reconstruct the IMF (variability). The contribution of low-mass stars to the spectrum is relatively small, and, as shown in DTK16, there is a degeneracy between the low-mass and the high-mass slopes. Since it is hard to distinguish dwarfs from giants in the wavelength range of the MILES library, we currently do not attempt more complicated parameterizations of the IMF prior. In future work, we plan to combine our model with the X-shooter Spectral Library \citep{XSL}. This empirical stellar library provides a good coverage of the HR diagram and has a wavelength range that extends from the UV to the NIR. The wavelength range of this library contains many more features that allow us to distinguish between dwarfs and giants. With an expanded wavelength range, we should also be able to probe more complicated parameterizations of the IMF prior.

\section*{Acknowledgements}
We thank the anonymous referee for the useful comments which improved the quality of this paper. This work was supported in part by NWO grant (project number 614.001.208) to SCT and by an NWO-VICI career grant (project number 639.043.308) to LVEK. RSS thanks the Downsbrough family for their generous support, and gratefully acknowledges support from the Simons Foundation through a Simons Investigator grant.




\bibliographystyle{mnras}
\bibliography{references} 


\appendix

\section{Binning isochrone stars}
\label{sec:AppIsochroneBinning}
To reduce the computational costs of running our model, we bin the stellar templates of an SSP. This procedure is discussed in Section \ref{sec:isochroneBinning}. When we combine different stellar templates into one new template, we need to assume an \textit{a priori} IMF, since binning affects the structure of matrix $\mathbfss{S}$ and cannot be done on-the-fly. In this paper we assume this a priori IMF in the binning procedure to have a Salpeter IMF slope inside the bin. This will slightly shift the average mass inside the bin. However, since the bins are very narrow, this assumption is of very low order. Here, we investigate the consequences of this assumption in more detail.

\begin{figure}
	\centering
		{\includegraphics[width=\columnwidth]{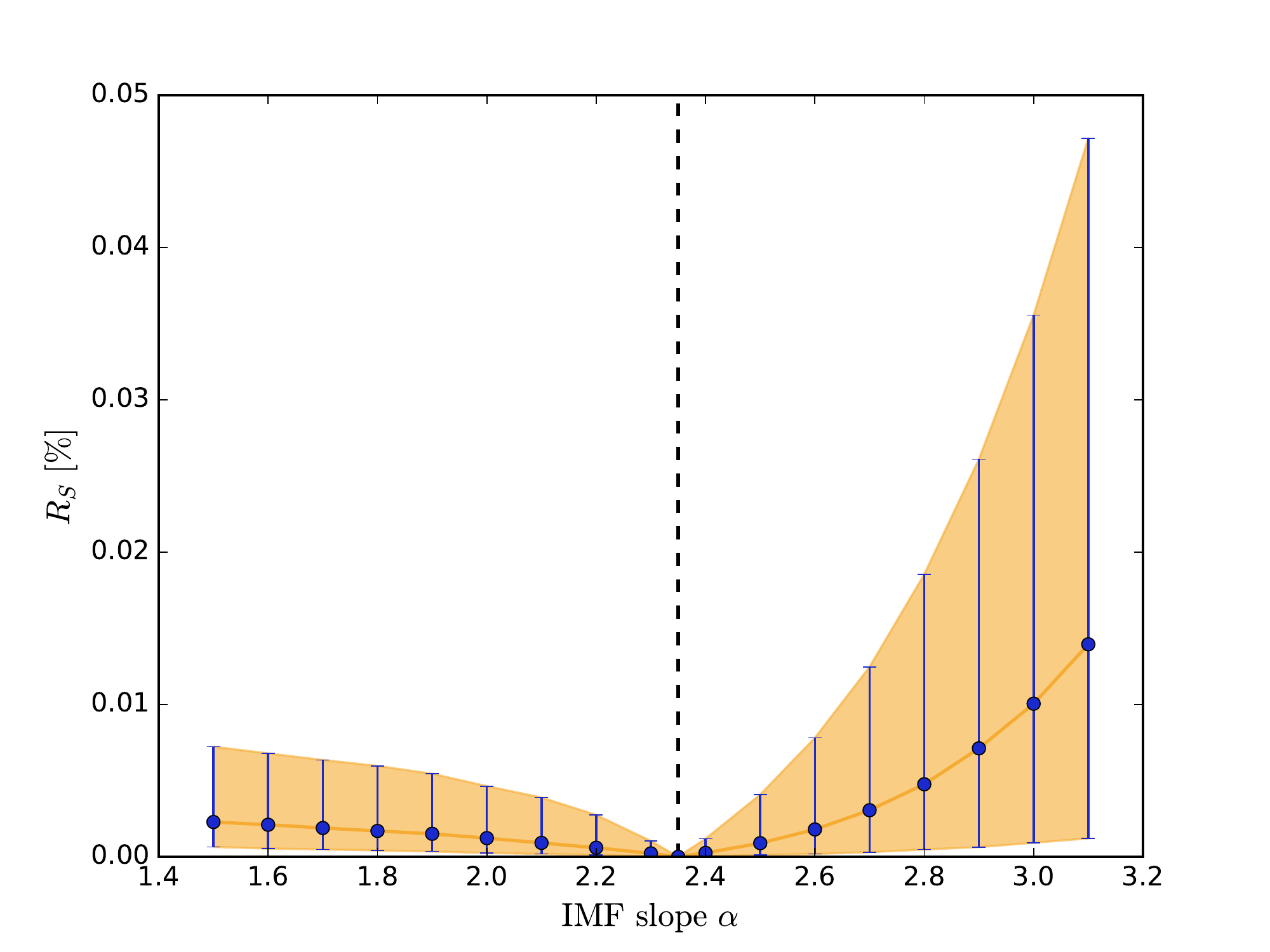}}
	\caption{Residuals between SSP spectra created with full set of stellar templates and SSP spectra created with binned version of stellar templates for different IMF slopes. The blue points correspond to the median residual of the age-metallicity grid for each of the IMF slopes that we consider. The error bars and the shaded orange region correspond to the 2$\sigma$ values of the distribution of residuals. The black-dashed line represent the Salpeter IMF slope.}
	\label{effectBinning}
\end{figure}

First, we determine the effect of our binning procedure on a set of SSP spectra. For every grid point in our age-metallicity grid, we create two SSP spectra: one created with the full set of stellar templates and one created with the binned set of stellar templates. Then, we calculate the residual between the two SSP spectra through
\begin{equation}
R_{\mathrm{S}} = \left\langle\frac{\mathrm{abs}\left( S_{\mathrm{SSP}} - S_{\mathrm{SSP,bin}} \right)}{S_{\mathrm{SSP}}}\right\rangle,
\end{equation}
where $S_{\mathrm{SSP}}$ is the SSP spectrum created with the full set of stellar templates and $S_{\mathrm{SSP,bin}}$ is the SSP spectrum created with the binned set of stellar templates. Assuming a single power law IMF, we repeat this for different IMF slopes in the range $\alpha = [1.5,3.1]$, where $\alpha=2.35$ corresponds to a Salpeter IMF. The results of this analysis are shown in Figure \ref{effectBinning}.  The residual between two SSP spectra is never more than 0.05\%. As expected, the residual becomes smaller as the IMF slope is closer to Salpeter and for a Salpeter IMF the two SSP spectra are exactly the same.

\begin{table}
	\centering
	\caption{Parameters of six mock SSPs that we use to test if binning the stellar templates under the assumption of a Salpeter IMF affects our inference of the IMF slope.}
	\label{tab:mockSSPsBinning1}
	\begin{tabular}{cccc} 
		\hline
		name & age [Gyr] & [Fe/H] & $\alpha$\\
		\hline
		\texttt{SSPt1} & 3.0 & 0.21 & 1.7\\
		\texttt{SSPt2} & 3.0 & 0.21 & 3.0\\
		\texttt{SSPt3} & 8.0 & -0.05 & 1.7\\
		\texttt{SSPt4} & 8.0 & -0.05 & 3.0\\
		\texttt{SSPt5} & 12.0 & -0.30 & 1.7\\
		\texttt{SSPt6} & 12.0 & -0.30 & 3.0\\		
		\hline
	\end{tabular}
\end{table}

Secondly, we test the effect of our binning procedure on the inference of the IMF. For this test, we have created six mock SSPs with different ages, metallicities and IMF slopes. These mock SSPs have been created with the full set of stellar templates, have an average SNR of 300 and are smoothed to a velocity dispersion of $150 \, \mathrm{km} \, \mathrm{s}^{-1}$. The parameters of the SSPs are summarized in Table A1. Then, we reconstruct the IMF slopes of these SSPs with our model using the binned version of the stellar templates. The reconstructed parameters are given in Table A2 and show that the inferred IMF slope for these SSPs is not affected by our binning procedure. Note that we considered two IMF slopes which significantly deviate from a Salpeter IMF slope, because if there is any effect, the effect is expected to be largest for these extreme slopes. Based on these tests, we conclude that the effect of binning can be fully neglected at the current levels of signal-to-noise (being $\ll 2000$ to reach the level of 0.05\% residuals).

\begin{table}
	\centering
	\caption{Reconstructed parameters of the mock SSPs in Table \ref{tab:mockSSPsBinning1}.}
	\label{tab:mockSSPsBinning2}
	\begin{tabular}{ccccc} 
		\hline
		{}  & \multicolumn{4}{c}{reconstructed} \\
		name & age [Gyr] & [Fe/H] & $\sigma$ & $\alpha$\\
		\hline
		\rule{0pt}{3ex}
		\texttt{SSPt1} & 3.0 & 0.21 & 150.3 & $1.66^{+0.04}_{-0.03}$\\
		\rule{0pt}{3ex}
		\texttt{SSPt2} & 3.0 & 0.21 & 150.5 & $3.03^{+0.02}_{-0.02}$\\
		\rule{0pt}{3ex}
		\texttt{SSPt3} & 8.0 & -0.05 & 150.3 & $1.70^{+0.04}_{-0.04}$\\
		\rule{0pt}{3ex}
		\texttt{SSPt4} & 8.0 & -0.05 & 149.8 & $2.99^{+0.02}_{-0.02}$\\
		\rule{0pt}{3ex}
		\texttt{SSPt5} & 12.0 & -0.30 & 150.2 & $1.71^{+0.03}_{-0.04}$\\
		\rule{0pt}{3ex}
		\texttt{SSPt6} & 12.0 & -0.30 & 149.9 & $3.00^{+0.01}_{-0.01}$\\		
		\hline
	\end{tabular}
\end{table}

\section{Systematic uncertainties}
\label{app:systematicUncertainties}
To get a feeling for the effect of systematic uncertainties on the differences in log evidence that we find, we perform an additional test with three mock SSPs. The ages and metallicities of these mock SSPs are given in Table \ref{tab:sysUncertainties}. All the mock SSPs in this Appendix have a Salpeter IMF and SNR = 150. First, we run the model with the complete set of stellar templates and calculate the evidence for the three mock SSPs. Then we remove the original stellar templates that were used to create the mock SSPs from the library of stellar templates (i.e. we remove all stellar templates with the age and metallicity of the mock SSP) and run the model again. In this way, we mock incompleteness of the basis set, which will give us some feeling on how to interpret differences in log evidence with respect to systematic uncertainties. 

The results of the additional test are summarized in Table \ref{tab:sysUncertainties}. For the youngest SSP, the difference in log evidence is relatively large. However, for the other two SSPs the difference in log evidence between running the model with and without the original stellar templates is well below ten. In all cases the model prefers to use the original set of stellar templates over the modified set of stellar templates. These results suggest that when we find a difference of log evidence that is more than $10^{3/2}\sim32$, which according to Jeffreys scale is considered as very strong evidence, that this difference in log evidence is meaningful in the context of systematic uncertainties. We emphasize though that the result of this test is also strongly dependent on the adopted spacings of the age-metallicity grid.
In Table \ref{tab:sysUncertainties_results} we show the inferred parameters of the SSPs that we obtain with the original stellar templates and with the modified stellar templates. The results that we infer with both sets of templates are consistent with each other within the errors, which implies that the results are robust against perturbations of the basis set of stellar templates.

\begin{table}
	\centering
	\caption{Evidence for three mock SSPs calculated with and without the original stellar templates that were used to create the mock SSPs. }
	\label{tab:sysUncertainties}
	\begin{tabular}{ccccc} 
		\hline
		name & age [Gyr] & [Fe/H] & evidence & evidence\\
		{} & {} & {} & with & without\\
		{} & {} & {} & original & original\\
		{} & {} & {} & templates & templates\\
		\hline
		\rule{0pt}{3ex}
		\texttt{SSPt7} & 3.0 & 0.21 & 10474.8 & 10450.5\\
		\rule{0pt}{3ex}
		\texttt{SSPt8} & 8.0 & -0.05 & 10506.4 & 10499.9\\
		\rule{0pt}{3ex}
		\texttt{SSPt9} & 12.0 & -0.30 & 10490.0 & 10486.5\\
		\hline
	\end{tabular}
\end{table}

\begin{table}
	\centering
	\caption{Reconstructed parameters of the three mock SSPs considered in Table \ref{tab:sysUncertainties} with the original set of stellar templates and with the modified set of stellar templates.}
	\label{tab:sysUncertainties_results}
	\begin{tabular}{cccc} 
		\hline
		name & reconstructed  & reconstructed & reconstructed\\
		{} & age [Gyr] & [Fe/H] & $\alpha$ \\
		\hline
		\multicolumn{4}{c}{original templates}\\
		\texttt{SSPt7} & {3.0} & {0.21} & {$2.33\pm0.04$}\\
		\texttt{SSPt8} & {8.0} & {-0.05} & {$2.33\pm0.04$}\\
		\texttt{SSPt9} & {12.0} & {-0.30} & {$2.33\pm0.03$}\\
		\multicolumn{4}{c}{modified templates}\\
		\texttt{SSPt7} & {2.8} & {0.21} & {$2.39\pm0.28$}\\
		\texttt{SSPt8} & {8.3} & {-0.05} & {$2.25\pm0.09$}\\
		\texttt{SSPt9} & {11.8} & {-0.30} & $2.34\pm0.05${}\\
		\hline
	\end{tabular}
\end{table}


\bsp	
\label{lastpage}
\end{document}